\documentclass{sigchi}

\CopyrightYear{2020}
\setcopyright{acmlicensed}
\doi{https://doi.org/10.1145/3313831.XXXXXXX}
\isbn{978-1-4503-6708-0/20/04}
\conferenceinfo{CHI'20,}{April  25--30, 2020, Honolulu, HI, USA}
\acmPrice{\$15.00}

\makeatletter
\def\@copyrightspace{\relax}
\makeatother

\usepackage{balance}       % to better equalize the last page
\usepackage{graphics}      % for EPS, load graphicx instead 
\usepackage[T1]{fontenc}   % for umlauts and other diaeresis
\usepackage{txfonts}
\usepackage{mathptmx}
\usepackage[pdflang={en-US},pdftex]{hyperref}
\usepackage{color}
\usepackage{booktabs}
\usepackage{textcomp}
\usepackage{multirow}
\usepackage[caption=false]{subfig}
\usepackage[table, dvipsnames]{xcolor}
\definecolor{Gray}{gray}{0.95}
\usepackage{newtxmath,tikz-cd}
\usetikzlibrary{decorations.pathreplacing,fit,shapes.geometric}
\usepackage{mathtools}
\usepackage[12pt]{moresize}
\usepackage{enumerate} 

\usepackage{longtable}

\definecolor{colora-block}{HTML}{4D79A8}
\definecolor{colora}{HTML}{B8C9DC}
\definecolor{colora-opaque}{HTML}{DBE4ED}

\definecolor{colorb-block}{HTML}{76B7B2}
\definecolor{colorb}{HTML}{C8E2E0}
\definecolor{colorb-opaque}{HTML}{E4F1F0}

\definecolor{colorc-block}{HTML}{F28E2B}
\definecolor{colorc}{HTML}{FAD2AB}
\definecolor{colorc-opaque}{HTML}{FCE8D5}

\definecolor{colord-block}{HTML}{E15759}
\definecolor{colord}{HTML}{F3BBBD}
\definecolor{colord-opaque}{HTML}{F9DDDE}

\usepackage{microtype}        % Improved Tracking and Kerning
\usepackage{ccicons}          % Cite your images correctly!
\usepackage{todonotes}

\newcommand{\etal}{et al.\@ }
\newcommand{\etals}{et al.\@'s }
\graphicspath{{figures/}{pictures/}{images/}{./}}

\usepackage{color}
\definecolor{mccolor}{rgb}{0.2,0.0,0.4}
\definecolor{amcolor}{rgb}{0.5,0,0}
\definecolor{glkcolor}{rgb}{0.5,0,0}

\ifx\hideComments\undefined
  \newcommand\mc[1]{{\color{mccolor}[MC: #1]}}
  \newcommand\am[1]{{\color{amcolor}[AM: #1]}}
  \newcommand\glk[1]{{\color{glkcolor}[GLK: #1]}}
\else
  \newcommand\mc[1]{{}}
  \newcommand\am[1]{{}}
  \newcommand\glk[1]{{}}
\fi

\newcommand{\figref}[1]{\hyperref[#1]{Figure~\ref*{#1}}}
\newcommand{\eqnref}[1]{\hyperref[#1]{Equation~\ref*{#1}}}

\def\plaintitle{Surfacing Visualization Mirages}

\def\emptyauthor{}
\def\plainkeywords{Information visualization; deceptive visualization; visualization testing}

\makeatletter
\def\url@leostyle{%
  \@ifundefined{selectfont}{
    \def\UrlFont{\sf}
  }{
    \def\UrlFont{\small\bf\ttfamily}
  }}
\makeatother
\def\pprw{8.5in}
\def\pprh{11in}

\definecolor{linkColor}{RGB}{6,125,233}
\hypersetup{%
  pdftitle={\plaintitle},
  pdfauthor={\emptyauthor},
  pdfkeywords={\plainkeywords},
  pdfdisplaydoctitle=true, % For Accessibility
  bookmarksnumbered,
  pdfstartview={FitH},
  colorlinks,
  citecolor=black,
  filecolor=black,
  linkcolor=black,
  urlcolor=linkColor,
  breaklinks=true,
  hypertexnames=false
}

\begin{document}

\title{\plaintitle}

\emptyauthor{}
\numberofauthors{3}

\author{
  \alignauthor{Andrew McNutt\\
    \affaddr{University of Chicago}\\
    \affaddr{Chicago, IL}\\
    \email{mcnutt@uchicago.edu}}\\
\alignauthor{Gordon Kindlmann\\
    \affaddr{University of Chicago}\\
    \affaddr{Chicago, IL}\\
    \email{glk@uchicago.edu}}\\
  \alignauthor{Michael Correll\\
    \affaddr{Tableau Research}\\
    \affaddr{Seattle, WA}\\
    \email{mcorrell@tableau.com}}
}

\maketitle

\begin{abstract}
% AM: can the abstract be longer than 150 words post accept? 
Dirty data and deceptive design practices can undermine, invert, or  invalidate the purported messages of charts and graphs.
These failures can arise silently: a conclusion derived from a particular visualization may look plausible unless the analyst looks closer and discovers an issue with the backing data, visual specification, or their own assumptions.
We term such silent but significant failures \emph{visualization mirages}.
We describe a conceptual model of mirages and show how they can be generated at every stage of the visual analytics process.
We adapt a methodology from software testing, \emph{metamorphic testing}, as a way of automatically surfacing potential mirages at the visual encoding stage of analysis through modifications to the underlying data and chart specification.
%AM ALT
% We adapt a methodology from software testing, \emph{metamorphic testing}, as a way of automatically surfacing potential mirages caused by algebraic errors between data and chart through modifications to the underlying data and chart specification.
We show that metamorphic testing can reliably identify mirages across a variety of chart types with relatively little prior knowledge of the data or the domain.

\end{abstract}

% ACM Classfication



% Author Keywords

% Print the classficiation codes
%Please use the 2012 Classifiers and see this link to embed them in the text: %\url{https://dl.acm.org/ccs/ccs_flat.cfm}

% AM Structural grammar questions
% - is it proof-of-concept or proof of concept, https://english.stackexchange.com/questions/124195/proof-of-concept-or-proof-of-concept-noun-or-adjective would seem to suggest the former?
% - how do you spell dataset? Is it data set or datasets?????
% - is it Author and OtherAuthor or Author & OtherAuthor? we do both right now which kind ssucks

\section{Introduction}
Visualizations, like all forms of communication, can mislead or misrepresent information. Visualizations often hide important details, downplay or fail to represent uncertainty, or interplay with complexities in the human perceptual system. Viewers often encounter charts produced by analytical pipelines that may not be robust to dirty data or statistical malpractice. It is straightforward to generate charts that, through deceit, accident, or carelessness, appear to show something of interest in a dataset, but do not in fact reliably communicate anything significant or replicable. We refer to the charts that superficially convey a particular message that is undermined by further scrutiny as visualization \emph{mirages}.

%In this paper, we present a conceptual model of these visualization ``mirages,'' show how they can lead to untrue or unwarranted conclusions from data, and present a prototype system that automatically parses the specification and backing data of a visualization to test for unreliable or deceptive properties. In addition to direct checks against known deceptive visualization practices our prototype also makes use of a concept called \emph{Metamorphic Testing}~\cite{segura2016survey} to explore an additional class of \emph{algebraic} errors~\cite{kindlmann2014algebraic}--- where minor changes to the visualization design or backing data have large (but illusory) effects on the resulting visualization, or where potentially important or disqualifying changes have no visual impact on the resulting visualization. Using our prototype, we show how metamorphic testing can identify a wide variety of important errors that would otherwise result in visualizations that encourage erroneous or unjustified conclusions.

In this paper, we present a conceptual model of these visualization mirages and show how users' choices can cause errors in all stages of the visual analytics (VA) process that can lead to untrue or unwarranted conclusions from data. Using our model we observe a gap in automatic techniques for validating visualizations, specifically in the relationship between data and chart specification. We address this gap by developing a theory of \textit{metamorphic testing for visualization} which synthesizes prior work on metamorphic testing~\cite{segura2016survey} and algebraic visualization errors~\cite{kindlmann2014algebraic}. 
Through this combination we seek to alert viewers to situations where minor changes to the visualization design or backing data have large (but illusory) effects on the resulting visualization, or where potentially important or disqualifying changes have no visual impact on the resulting visualization. We develop a proof of concept system that demonstrates the validity of this approach, and call for further study in mixed-initiative visualization verification.

%We highlight our contributions as being three-fold.

%\begin{enumerate}
%    \item We introduce the terminology of \emph{visualization mirage} and show how these mirages can occur at all stages of the analytics process.
%    \item We introduce \emph{metamorphic testing} as a framework for automatically surfacing potential data of visual design issues that can lead to mirages.
%\end{enumerate}

% Code and related materials are available at \mc{GITHUB/OSF LINK}. 

\begin{figure}[bth]
   \centering
   \includegraphics[width=\columnwidth]{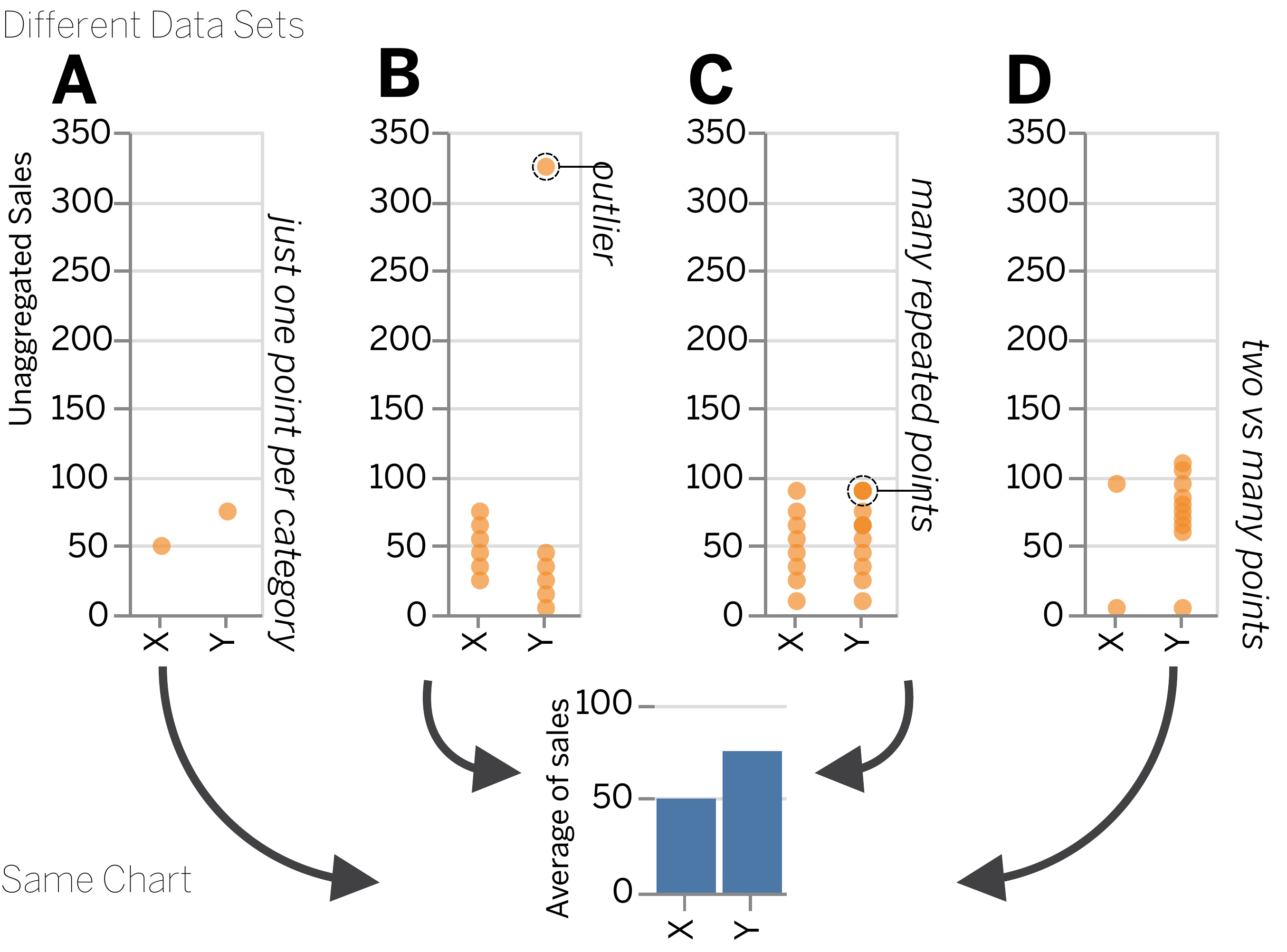}
   \caption{
   Visualizations that aggregate data can mask their underlying data distributions, which can create mirages, as in this bar chart over synthetic data. Not all distributions equally support the message that sales in Y are reliably higher than X. For instance, this difference could be driven by a single outlier (B), a possibly erroneously repeated value (C), or be an artifact of high variability caused by low sample size (D).
   }
   \label{fig:deception-quartet}
      \vspace{-0.05in}
\end{figure}

\section{Visualization Mirages}

We define a visualization mirage as follows: 

\begin{quote}
A \textbf{visualization mirage} is any visualization where the cursory reading of the visualization would appear to support a particular message arising from the data, but where a closer re-examination of the visualization, backing data, or analytical process would invalidate or cast significant doubt on this support.
\end{quote}

A long sequence of contexts and decisions, from the initial data curation and collection, to the eventual reader's literacies and assumptions, determine the message that a visualization delivers.
%Sacha \etal~\cite{sacha2015role} explore how trust and uncertainty accrete at different stages of this anaylsis process. We, in turn, look at the creation of falsehood.
Mistakes, errors, or intentionally deceptive choices anywhere along this process can create visualization mirages, from dirty data~\cite{kim2003taxonomy} to cognitive biases~\cite{dimara2018task}. Failures can occur at an early stage, but not result in a mirage until a later stage. 
For instance, missing data as a result of an error in data collection may be visible in a particular visualization, such as univariate data in a dot plot, and so be unlikely to lead to an error in judgment. Yet this data flaw may be invisible in a less robust visualization-design such as a histogram~\cite{correll2018looks}. Whether the missing data results in a mirage is contingent on the choice of eventual visualization design.

Mirages also depend on the reader's task. What may be misleading in the context of one task may not interfere with another. For instance, bias in estimating angle caused by a poorly selected aspect ratio~\cite{heer2006multi} or improperly selected scales~\cite{cleveland1982variables} could potentially produce a mirage for a viewer interested in correlation, but is unlikely to impact a viewer concerned with locating extrema. \figref{fig:deception-quartet} shows how mirages can arise for the task of comparing values in a bar chart: while the final bar chart is identical for all for cases A-D, some of these cases suggest statistical or data quality concerns that would cause a reader to doubt the reliability or robustness of any conclusions drawn from the direct comparison of values in the bar chart.
%Detecting these mirages, or employing designs that are robust against failures of this sort, can lead to better decision-making and encourage both data literacy and data skepticism in our audiences. \am{Citation, or is this a claim we're making?} 

%\am{Every time i go through this i feel like we should be citing either jock or hibbard about effect. maybe: Charting errors that misalign the types of data with the chart type (such as using pie charts to describe a temporal variable) are not necessarily mirage} \am{And everytune I go to fix, i have no idea what i was referring to, blergh}
Not all errors in charting are mirages; for a viewer to be mistaken, the visualization must appear to credibly communicate \emph{something}. 
Errors that fail to generate visualizations (such as software errors), or generate errors that are readily visible (e.g. ``glitch-charts''~\cite{glitchTumblr}) do not function as mirages: either there is no illusory message to dispel, or the visualization has visible irregularities that cast its accuracy into immediate doubt. 
We also exclude charts which are intended to be consumed only as art, that Ziemkiewicz \etal\cite{ziemkiewicz2009embedding} place outside of the traditional rhetorics of information visualization, although many artistic or ambient visualizations can and do persuade~\cite{moere2007towards}, and so have the potential to mislead.
We instead focus on cases where the chart appears correct, but where tacit issues in the analysis and generation pipeline have produced a mismatch between the conclusions supported by the data and the message communicated by the chart.

Mirages pose an important design problem: tools in the VA process should help to augment~\cite{heer2019agency} and enhance the reader's understanding of their data in such a way that the reader either automatically avoids mirages or is alerted to them in a useful manner. Simply alerting the reader to potential issues in a given visualization may be sufficient to provoke skepticism or follow-up analysis that would dispel the mirage.

%We enumerate choices that can produce mirages and the categories of errors that they can create in \figref{fig:mirage-figure}. 

% A Table of Contents paragraph oh boy
In the next section we provide a narrative exemplifying how mirages can arise in the course of an analysis session. We then provide an in-depth discussion of the ways that mirages can form and the problems they cause. We then consider prior systems for addressing these errors, and introduce the use of our \textit{Metamoprhic Testing for Visualization} technique for identifying visual encoding-based mirages. We substantiate our technique through a computational experiment.
\begin{figure}[h!]
   \centering
   \includegraphics[width=0.9\columnwidth]{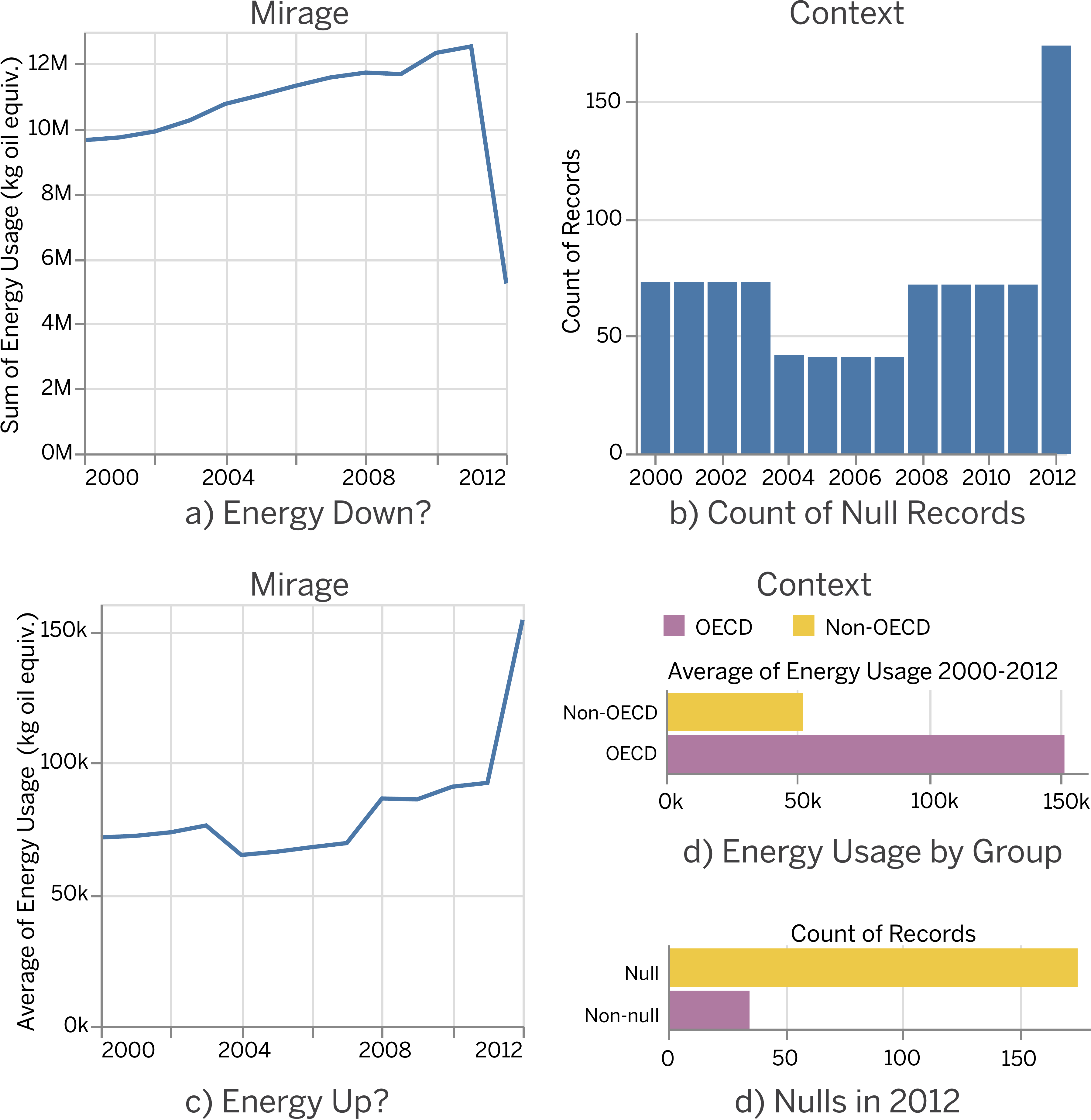}
   \caption{
   Do we always see what we get? 
   Here we show energy usage by countries from the ``World Indicators" dataset \protect\cite{worldBank}.
   %   Here we visualize energy usage by countries from the ``World Indicators" dataset \protect\cite{worldBank} in service of the question ``What is the trend of global energy usage over time, and what drives this trend?"
   Even these simple charts contain subtleties that can cause a careless analyst to make errors that give way to mirages. For instance, the choice of aggregate can make the trend in global energy usage appear to be either dramatically increasing or decreasing in 2012. But neither line chart reveals the large number of missing values in 2012, or the biased nature of the values that remain.
   }
   \label{fig:world-indicators-up-down}
      \vspace{-0.25in}
\end{figure}

\subsection{Illustration: World Indicators}

Mirages arise in the ordinary process of visual analytics, and can require significant effort or knowledge to detect and diagnose.
We present a sample VA session and highlight mirages based on real situations encountered by the authors. 
Through this analysis we aim to answer a question: ``What is the trend of global energy usage over time and what drives this trend?"
We focus on the ``World Indicators" dataset~\cite{worldBank} which consists of per-country  statistics from 2000-2012. We choose this dataset because of its prior use in showing the value of visualization to understand the trajectory of human development~\cite{rosling2011health}.

We begin our analysis by constructing a time series visualization of energy usage. 
\figref{fig:world-indicators-up-down}a appears to show a sharp decrease in energy usage in 2012, which may indicate a worldwide shift towards sustainability initiatives. However, this decrease is illusory, and is caused by a problem in curation: there are far fewer records for 2012 than in previous years, as shown in \figref{fig:world-indicators-up-down}b.
To lessen the impact of these missing values, we change from aggregating by SUM to aggregating by MEAN. The resulting visualization, \figref{fig:world-indicators-up-down}c, now appears to show the \emph{opposite} trend: energy usage sharply increased in 2012. However, the missing records combined with our choice of how to wrangle the data has created another mirage. The only non-null entries for 2012 are OECD (Organisation for Economic Co-operation and Development) countries. These countries have significantly higher energy usage than other countries across all years  (\figref{fig:world-indicators-up-down}d).

\begin{figure*}[t!]
   \centering
   \includegraphics[width=2\columnwidth]{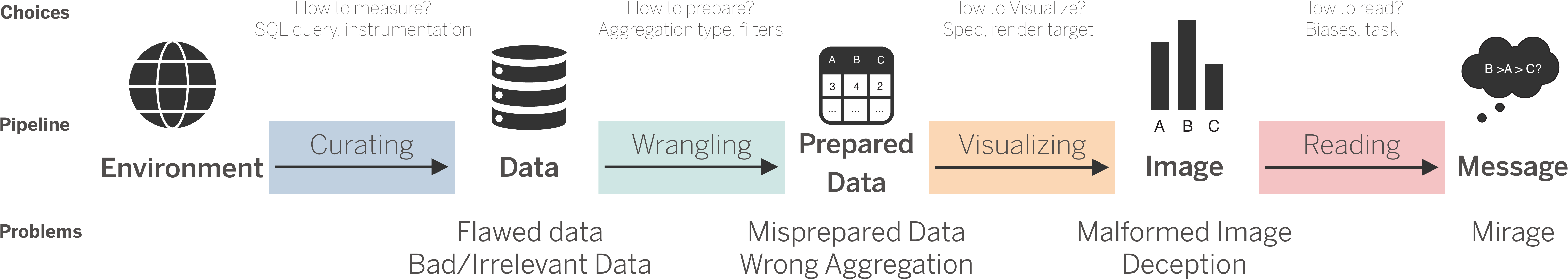}
   \caption{
   Each stage in the VA process offers users agency but also presents an opportunity for errors, which can give way to mirages.
    We see errors as occurring at the discrete stages of our pipeline (symbols), which are caused by the  choices made by users (arrows). 
   This framing allows us to identify the dependencies between actions, errors they create, and ways that those errors can propagate to deceive the reader.
   }
   \label{fig:mirage-figure}
            \vspace{-0.15in}
\end{figure*}

\begin{figure}[h!]
   \centering
   \includegraphics[width=0.9\columnwidth]{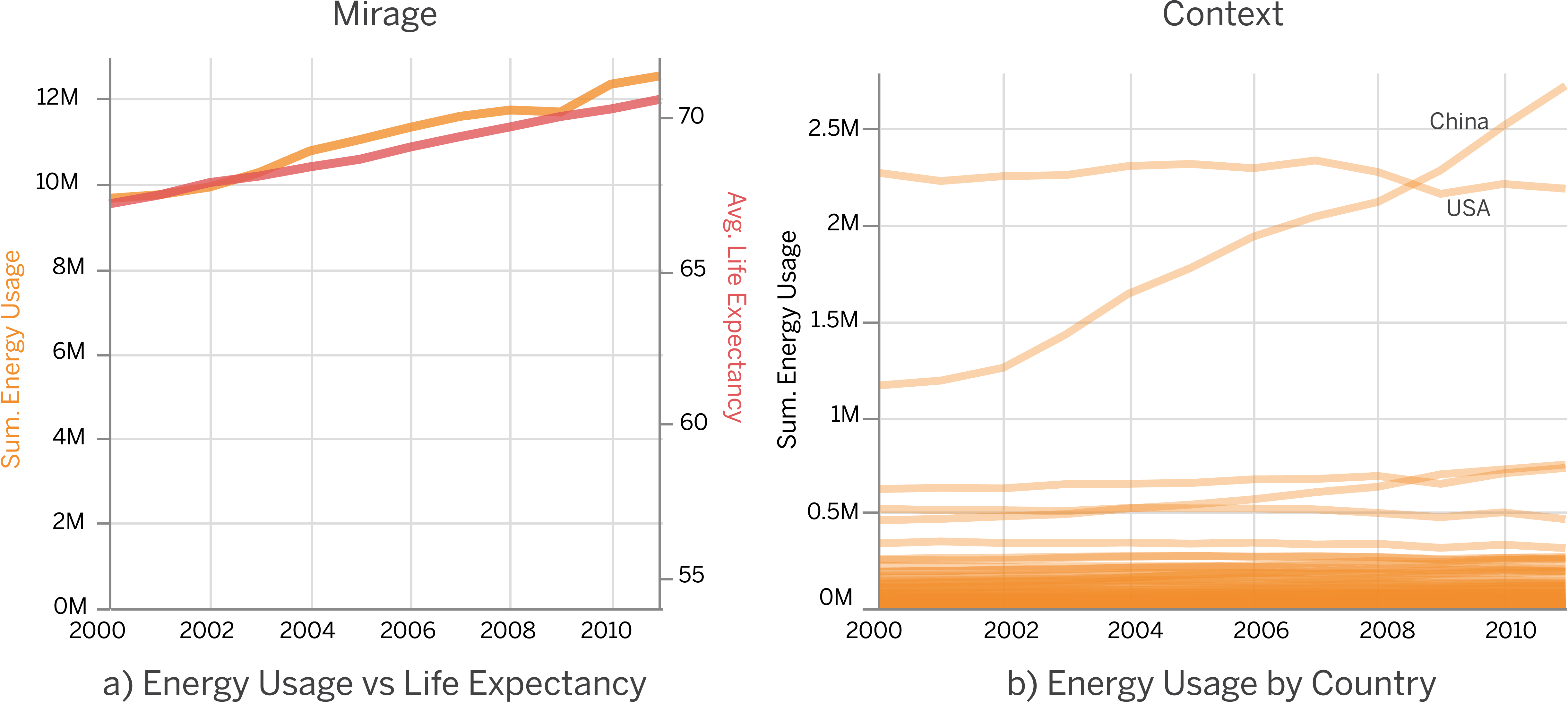}
   \caption{
   Energy usage compared to life expectancy in the ``World Indicators" dataset \protect\cite{worldBank} appear to be well correlated, but that inference is an illusion driven by two significant outliers (the US and China) and by a malicious manipulation of one of the y-axes (in particular starting the right y-axis from 55 instead of 0). 
   }
   \label{fig:world-indicators-dual}
         \vspace{-0.15in}
\end{figure}

Given these irregularities, we filter out 2012 and focus instead on the gradual upwards trend in energy usage for the remaining years. This upwards trend might indicate energy usage corresponds with a general increase in the length and quality of life. \figref{fig:world-indicators-dual}a appears to support this hypothesis: worldwide energy usage appears to be tightly correlated with life expectancy. This, too, is a mirage. The y-axis of the chart has been altered to make the rates of increase appear similar: while average life expectancy only increased 4\% from 2000-2011, overall energy usage increased 30\%. These y-axis manipulations can bias viewers of the chart~\cite{cleveland1982variables,pandey2015deceptive}, as can plotting two potentially unrelated variables in the same chart~\cite{xiong2019illusion}. We would categorize the latter mirage, arising from the Reading stage of our pipeline, to as \textit{Assumptions of Causality} (see Table 2).

%It easy to assume that the dataset that you have access to fully captures the space being addressed. Assumptions of fal
%This analysis overlooks part-to-whole charts because they were not presented as the default views by the VA under use in this context (Tableau).

Moreover, aggregating all countries together obscures considerable variability in the purported universal trend. When we disaggregate the data and remove the dual-axis, we see that much of the global growth in energy usage is attributed to China (whose usage more than doubled across the time period in question), moderated by a slight decrease in energy usage in the United States \figref{fig:world-indicators-dual}b. These countries dominate the trend, with most of the remaining countries having relatively flat trends when plotted in the same visual space.

%While changes to data model and visualization design that can dispel some false or ephemeral impressions, many visualization fail to present any indication that something odd is going on. 
In the absence of automated or semi-automated tools to highlight potential concerns, it is up to the attention, skepticism, domain knowledge, and statistical sophistication of the consumer of the visualization to attempt to verify the accuracy of what they are seeing. The visualizations themselves fail to provide any indication of potential errors of interpretation, and indeed many appear to present a clear, reasonable, and final answer to the questions posed by the analyst. 

%\begin{figure*}[t!]
%   \centering
%   \includegraphics[width=2\columnwidth]{mirages.pdf}
%   \caption{
%   Each stage in the visual analytics process offers users agency but also presents an opportunity for errors, which can give way to mirages.
   %Visualization mirages can come from all points of action in the visualization pipeline.
   %
   %While our model does not reflect the iterative loops and cycles of the analysis and sense-making process, it allows us to identify the dependencies between actions, errors they create, and ways that those errors can propagate to deceive the reader.
   %
   %We differentiate our model from others in that we do not try to explicitly model each step and loop in the process, instead favoring a unidirectional model (reminiscent of Chen \etals information theory-based model of visualization\protect\cite{chen2010information}).
   %
 %  This allows us to identify the dependencies between actions, errors they create, and ways that those errors can propagate to deceive the reader.
   %We use colored blocks throughout the paper to indicate issues that can arise during the Curation \CURATING, Wrangling, Visualizing, and Reading of data.
 %  }
%   \label{fig:mirage-figure}
%\end{figure*}

\section{Where Do Mirages Come From?}

We show how choices can create errors and highlight the way that those errors can propagate to become mirages in \figref{fig:mirage-figure}. 
Following Heer \cite{heer2019Visualization}, we focus on moments of agency in the visual analytic process (denoted in our diagram by arrows) that can introduce failures either by themselves, or in concert with other decisions, to generate visualization mirages. 
%This perspective gives us clues as to the causal relationship between choices and mirages, which is beneficial to our larger goal of being able to automatically surface these issues to the chart creator. 
This perspective gives us clues to the causal relationship between choices and mirages, which is beneficial for automatically surfacing these issues to the chart creator. 

Our model builds upon pipeline-based descriptions of problems in the VA process~\cite{chi2000taxonomy, pirolli2005sensemaking, van2005value, vickers2012understanding}. Our work most notably expands upon Borland \etals \cite{borland2018contextual} categorization of where threats to validity arise in VA, Sacha \etals\cite{sacha2015role} model of how visualization awareness and trust disseminate across the sensemaking loops of the process, and Heer's \cite{heer2019Visualization} description of the points of failure across the analytics process. While we acknowledge that real-world analytics processes include many cycles and nested sub-processes~\cite{pirolli2005sensemaking}, our simplified pipeline allows us to directly \emph{attribute} errors to specific points, and \emph{trace} those errors to the resulting mirage. 
%Different parts of this pipeline are more or less amenable to automatic detection or correction. For instance, errors in reading may require statistical education or critical reflection, whereas errors in data quality or statistical analysis could be automatically surfaced.
Errors in some parts of this pipeline are more amenable to automation than others. For instance, errors in reading may require statistical education or critical reflection, whereas errors in data quality or statistical analysis could be automatically surfaced.

\subsection{Data-Driven Mirages}

\begin{figure}[bth]
   \centering
   \includegraphics[width=0.6\columnwidth]{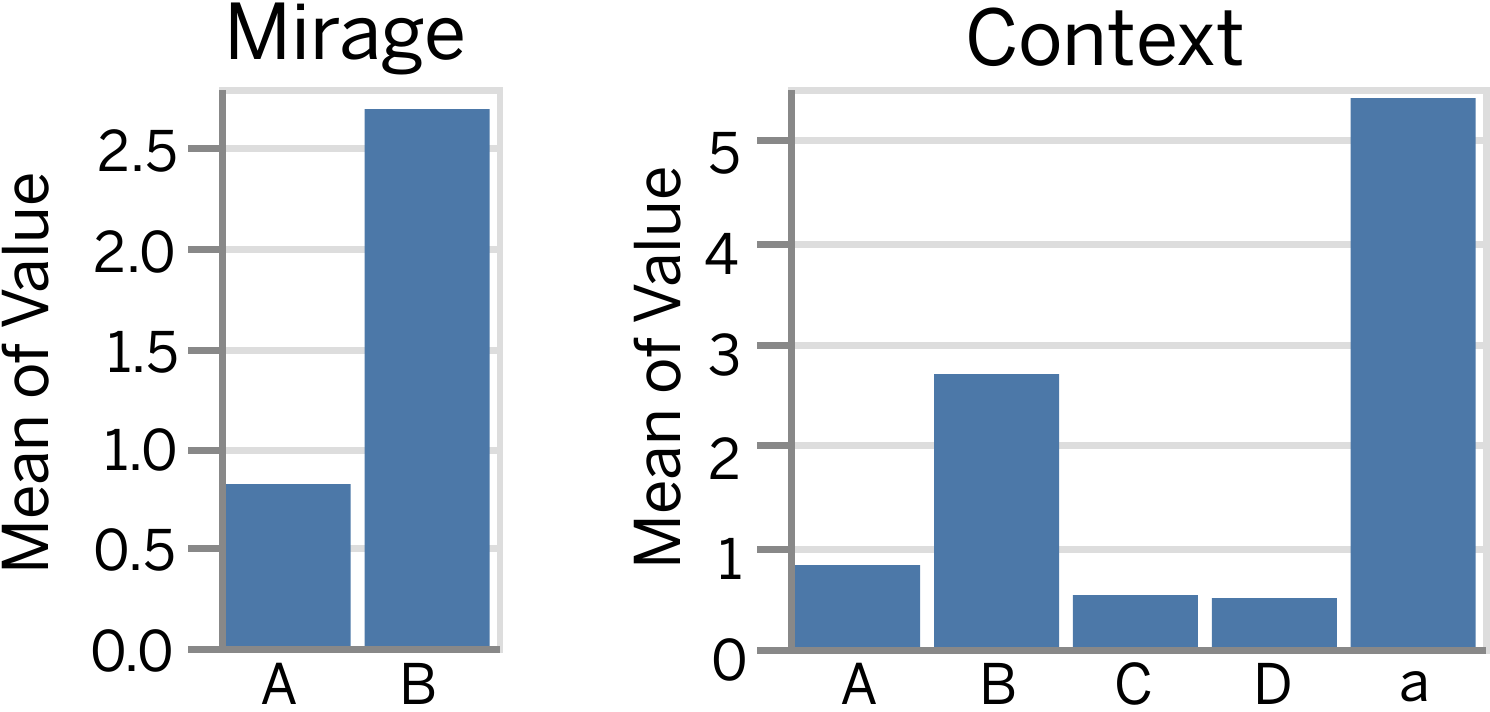}
 
   \caption{
      Data curation errors can cause mirages. The first chart uses a case-sensitive filter to directly compare `A' and `B', hiding that fact that many values were mistakenly entered as `a' and producing the impression that sales for `B' are significantly higher than `A'. If these `a' values are merged with `A,' this apparent difference would reverse.
%   Mirages can be caused by data curation errors, such as in this synthetic example in which a spelling mistake in the curation process causes a grouping error, which in turn causes an inaccurate inference about the data.
   }
   \label{fig:misspelling}
\end{figure}

\begin{figure}[bth]
   \centering
   \includegraphics[width=0.9\columnwidth]{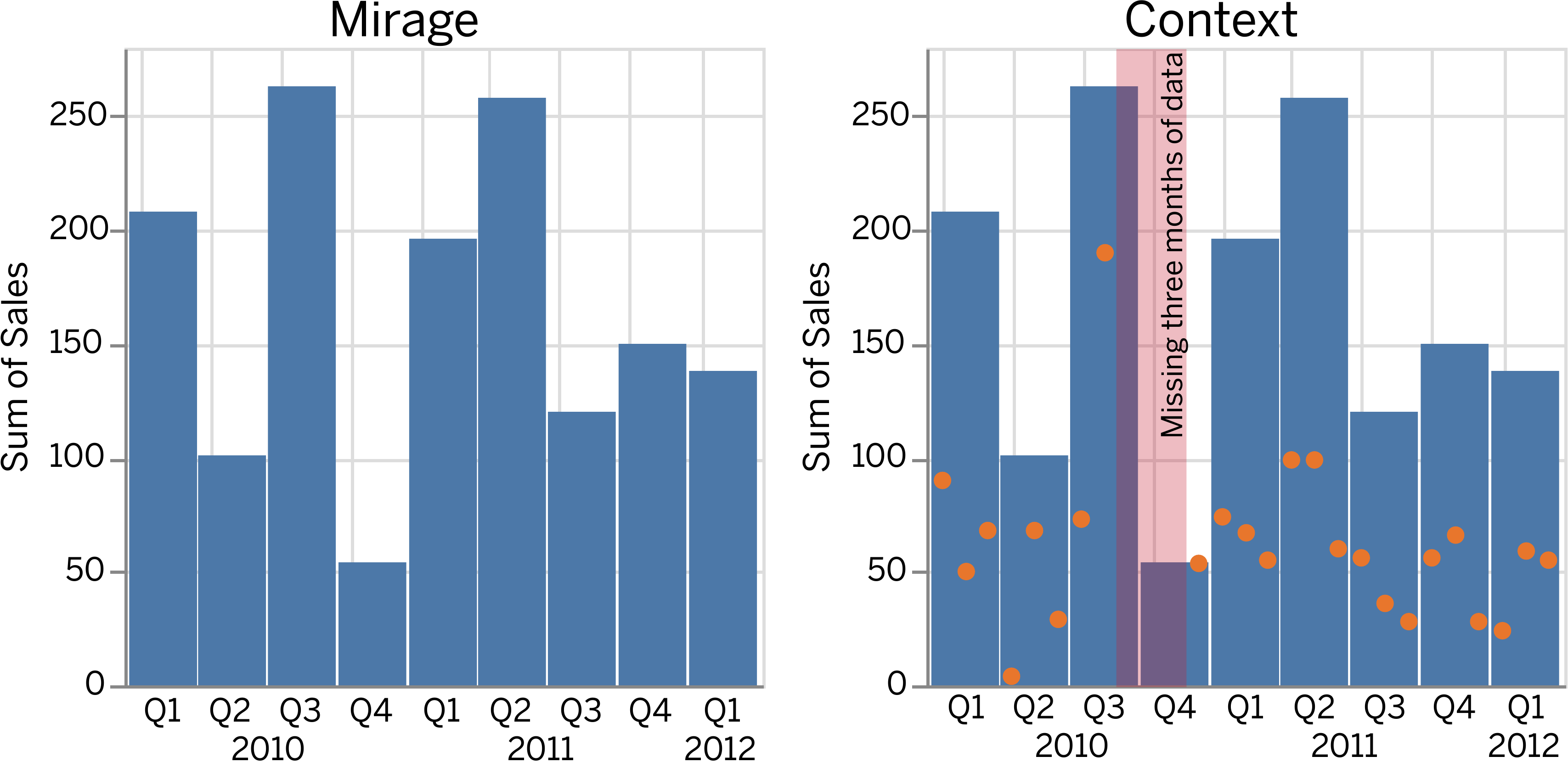}
   \caption{
   Wrangling and curation errors can cause mirages. In this synthetic example, the decision to aggregate hides three months of missing data, and gives the impression that sales were down in 2010Q4.
   }
   \label{fig:wrangling}
\end{figure}

We use the term \textbf{Curating} to denote the entire process of collecting, measuring, organizing, and combining data. Once the datasets are created, the analyst must clean, filter, subset, model, and shape the data into a form that is usable by the visualization system. We refer to this step as \textbf{Wrangling}. 
The efficacy of a visualization is limited by the quality and characteristics of its backing data. Even the most well-designed chart will be fundamentally flawed if the data on which it relies is irrelevant, incomplete, or biased, or has been processed or combined carelessly. The resulting ``dirty data'' (see Kim \etal\cite{kim2003taxonomy} for a taxonomic overview) can lead to mirages as in \figref{fig:misspelling} and \figref{fig:wrangling}.
Tang \etal\cite{tang2019towards} describe the challenge of automatically detecting and understanding the ways in which dirty data can create misleading trends as one of the most important open problems in visualization.

\subsection{Design-Driven Mirages}
%\am{A lof of this section is just listed charting problems, but not necessarily mirages, or at least they aren't prhased as mirages}

\begin{figure}[t]
   \centering
   \includegraphics[width=.8\columnwidth]{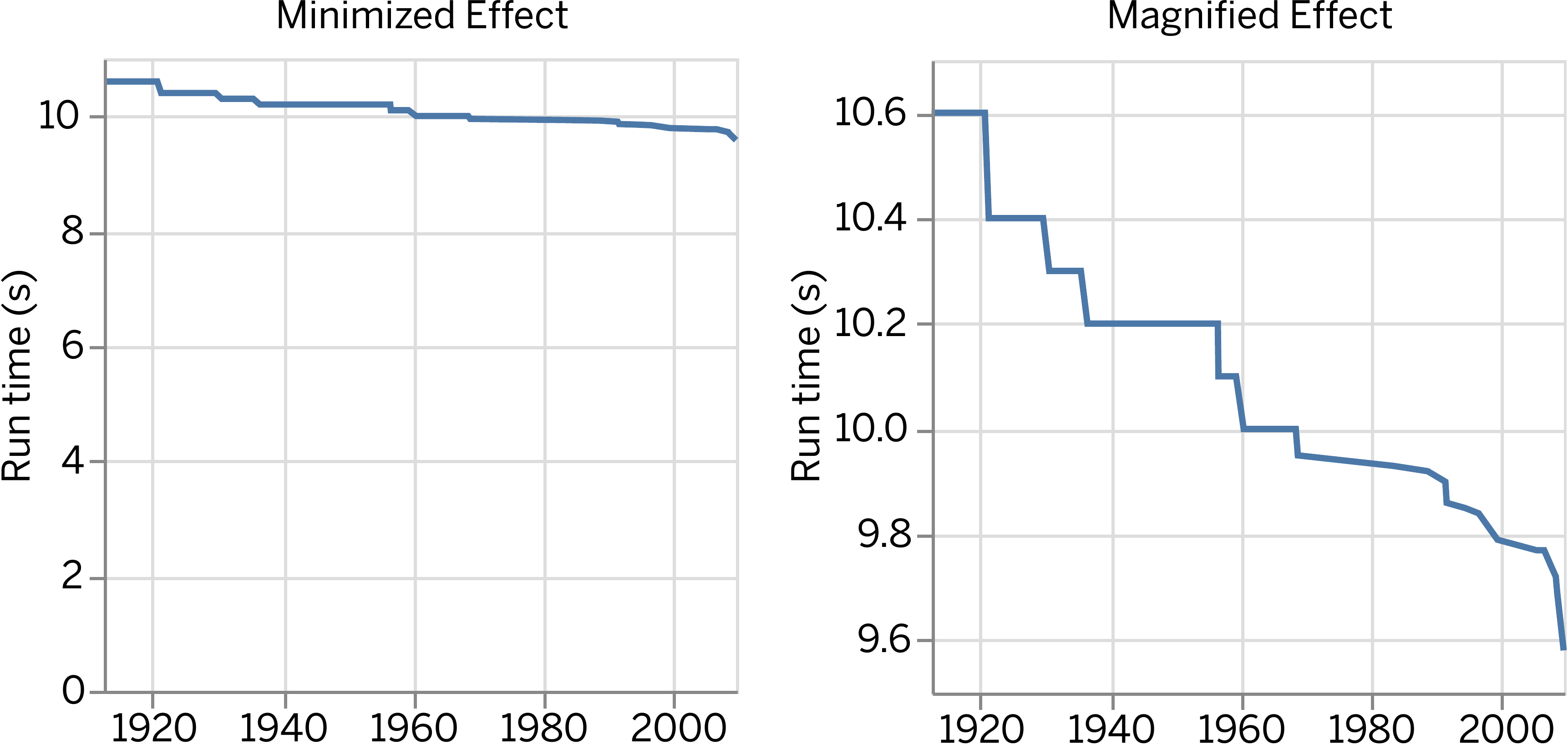}
    
   \caption{
  Visualizations can be grammatical yet deceptive. Changes to Men's 100-meter dash world record over the last century. The left image suggests that there hasn't been much change in that time, yet the image on the right shows a consistent pattern of improvement. The relevant effect size (is a 1 second improvement important or not?) can be exaggerated or hidden by the designer's choice of axis.
   }
   \label{fig:deceptive-axes}
\end{figure}

Once the data are in a proper form, the next step is to render the data in some human-legible way. We refer to this step as \textbf{Visualizing}. The last step is then for a human to read, interpret, and comprehend the resulting design. This \textbf{Reading} involves the literal decoding of the visual variables in a chart to retrieve the values of interest, as well as reasoning about the importance of the patterns identified in the data and updating prior beliefs based on new information.

Many visualization designs are known to be deceptive or prone to misinterpretation, as in \figref{fig:deceptive-axes}~\cite{bresciani2009risks,bresciani2015pitfalls,cairo2015graphics,cairo2019,correll2017black,huff1993lie,szafir2018good,wainer1984display}. Pandey \etal\cite{pandey2015deceptive} find that commonly discussed errors such as truncated y-axes and size/area confounds impact subjectively assessed trends and differences in values. Kong \etal\cite{kong2018frames,kong2019trust} find that slanted and biased chart titles can impact how the data are later recalled. Cleveland \etal\cite{cleveland1982variables} find that scale alterations can bias the perception of correlation in scatterplots, and Newman \& Scholl~\cite{newman2012bar} find that bar charts create a bias when viewers estimate the likelihood of samples. These deceptive practices, and the biases they induce, can create mirages. In addition, prior knowledge or priming can result in viewers having different interpretations of the same data~\cite{xiong2019curse}. Inattention to accessibility may also create mirages. For instance, designers that are not mindful of color vision deficiencies can create visualizations that communicate markedly different messages to different audiences~\cite{plaisant2005information}.

%There is a wide class of visualization designs that are known to be deceptive or prone to misinterpretation, as in \figref{fig:deceptive-axes}. Bresciani \& Eppler~\cite{bresciani2009risks,bresciani2015pitfalls} enumerate various ``risks'' and ``pitfalls'' that can occur when interpreting visualizations. Huff~\cite{huff1993lie} presents many examples of misleading charts in \emph{How to Lie With Statistics}, Wainer~\cite{wainer1984display} likewise presents a guide on ``how to display data badly,'' and Szafir \etal~\cite{szafir2018good} enumerate ``five ways that visualizations can mislead.'' 
%More recently, Correll \& Heer~\cite{correll2017black} enumerate a class of visualization ``attacks'' that do not misrepresent the data \emph{per se} but may still mislead the viewer, and Cairo has explored numerous ways that charts can ``lie''~\cite{cairo2015graphics,cairo2019}. \am{Should add some examples of mirages from this category}

\subsection{Mirages at the Intersection of Data and Design}

%\am{Add a motivating section here, this is the setup to metamorphic testing, and it needs to be sung loud and proud}
%- Talk about "looks good to me" \cite{correll2018looks}
%- avd \cite{kindlmann2014algebraic, KindlmannAlgebraicVisPedagogyPDV2016}
%- demerilap perhaps \cite{demiralp2014learning}
%- hibbard and jock \cite{hibbard1994lattice} \cite{mackinlay1986automating}
%Visual

The data and visualization may not have problems on their own, yet still create mirages when combined. For instance, Correll \etal\cite{correll2018looks} describe how data errors, including outliers and missing values, may fail to be detectable in univariate visualizations only for certain design parameters.
%Their work grows out of Kindlmann \& Scheidegger's \cite{kindlmann2014algebraic, KindlmannAlgebraicVisPedagogyPDV2016} algebraic model of the ways that visualizations can fail to be robust to data changes. Hibbard \etal\cite{hibbard1994lattice} provides a similar model by formalizing Mackinlay's effectiveness principles \cite{mackinlay1986automating}.
%Pu \& Kay~\cite{pu2018garden} caution that visual analytics systems that allow arbitrary amounts of filtering, drill-down, and sampling, can function as ``p-hacking machines'' that generate spurious conclusions. 
Zgraggen \etal\cite{zgraggen2018investigating} found that, in systems without visualizations of statistical uncertainty or control for robustness, many ``insights'' reported from a sample of a dataset failed to be true of the larger dataset. 
In Simpson's paradox~\cite{armstrong2014visualizing,guo2017you} patterns can appear to reverse based on the level of aggregation. Lee \etal\cite{lee2019avoiding} describe the drill-down fallacy wherein ignoring explanatory variables during the process of filtering can result in erroneous claims of causality.
There has been relatively little scholarship formalizing the problems that can occur in the specific relationship between data and chart \cite{hibbard1994lattice, kindlmann2014algebraic, KindlmannAlgebraicVisPedagogyPDV2016, mackinlay1986automating}, which entails a corresponding gap in testing strategies for automatically probing for problems.

\begin{figure}[t]
   \centering
   \vspace{-0.15in}% trying to get this figure to be aligned with the one to it's right at the top of the page
\subfloat[\small{Trulia crime map~\protect\cite{trulia}}]{\includegraphics[width=.41\linewidth]{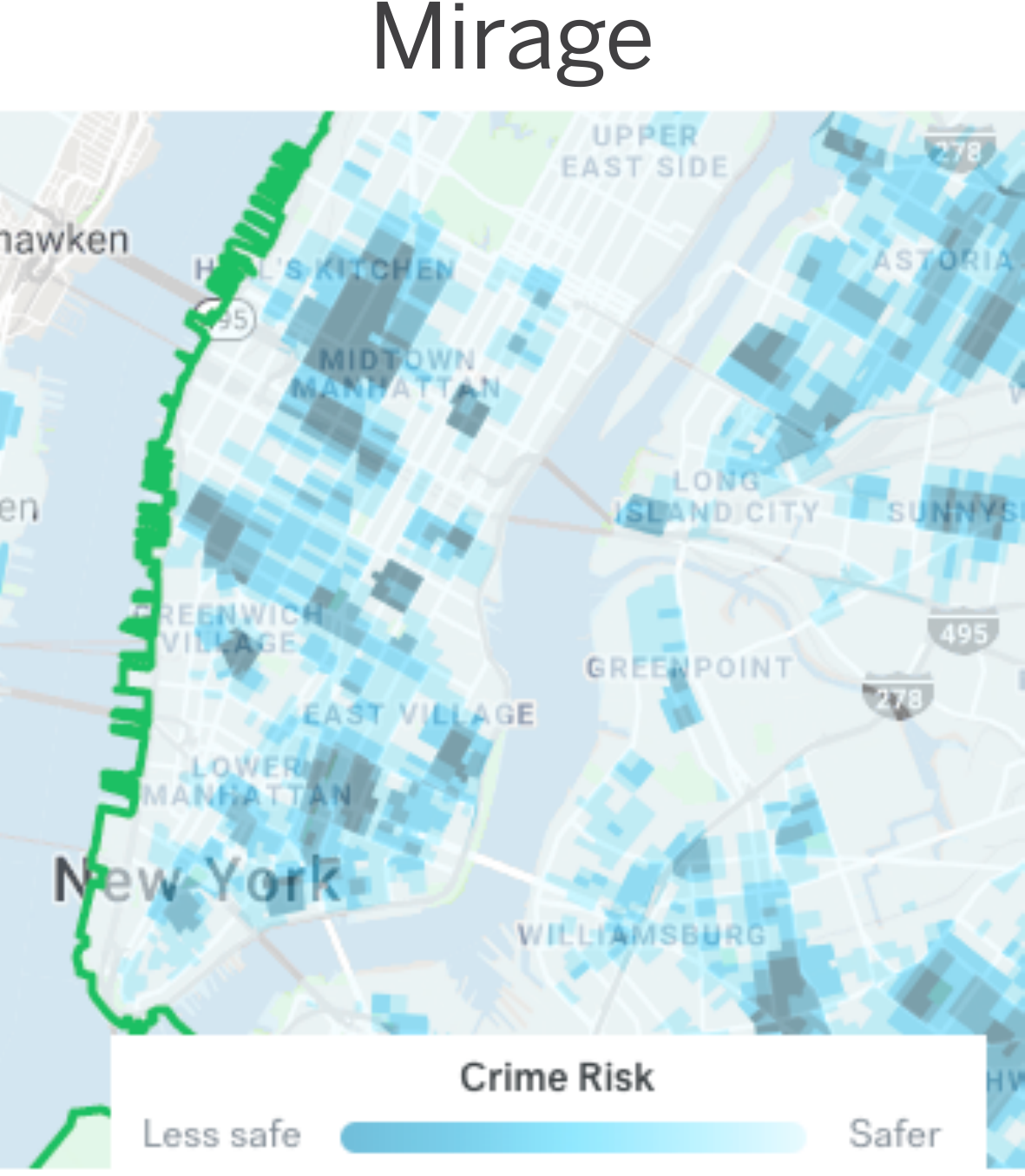}}\qquad
\subfloat[\small{White collar crime map \protect\cite{lavigne2017predicting}}]{\includegraphics[width=.41\linewidth]{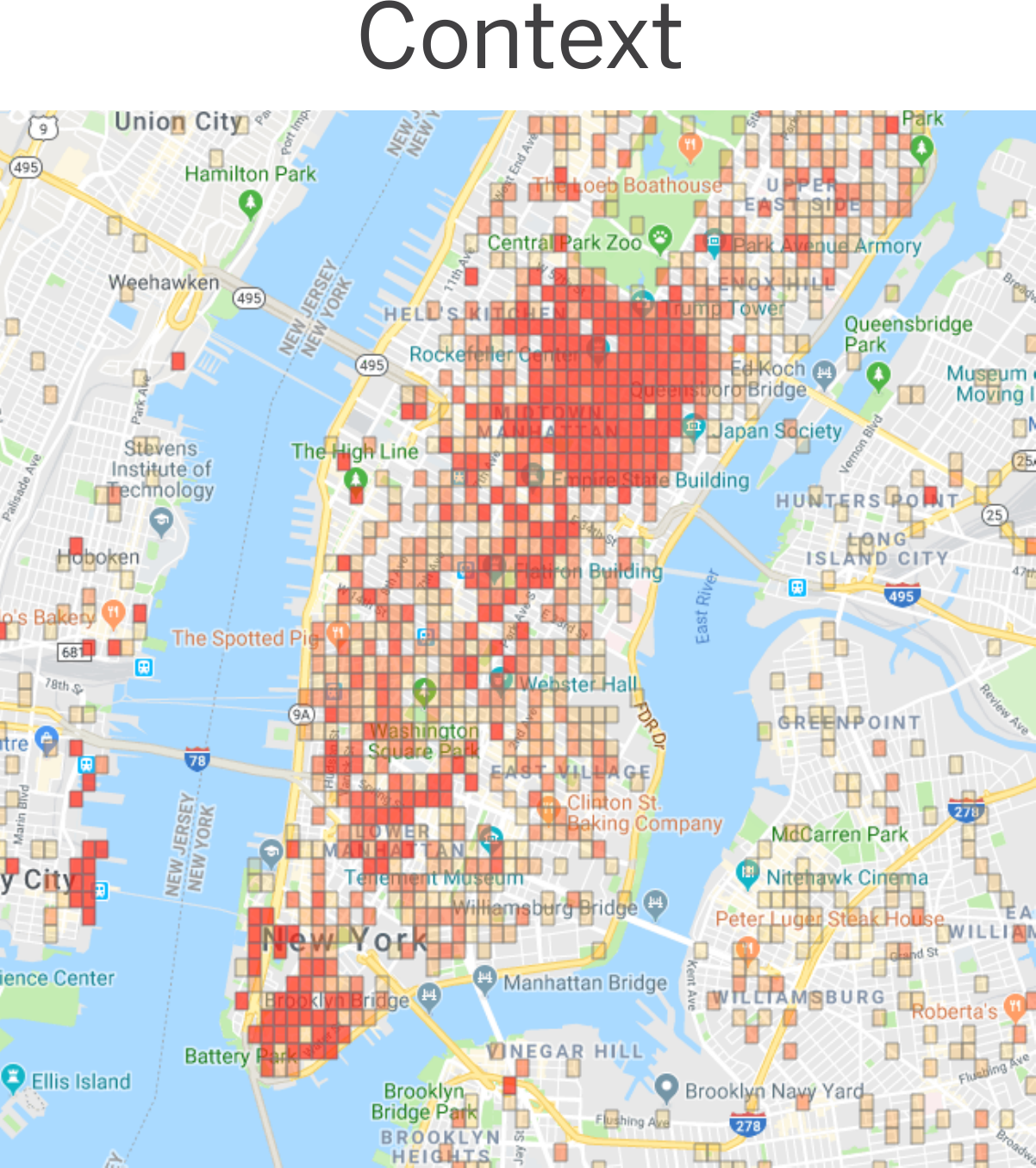}}
   \caption{Crime maps typically are generated by 911 calls and police reports, which capture only a subset of criminal activity.  White collar and financial crimes are often excluded from these maps; their inclusion paints a different picture of locations in the city where the most crimes occur. The choice of which data to include or exclude, and biases in how these data are collected, can control the message of the chart.
   }
   \label{fig:missing-datasets}
\end{figure}

\subsection{Other Sources of Mirages}

Reading errors can occur in conjunction with other parts of the pipeline, such as with Curation errors.
For instance, biases and assumptions on the part of the chart creators and readers can skew the resulting messages gleaned from charts \cite{dimara2018task, valdez2017framework}. 
% \am{feels like were missing a cite for persuasive/subjective charts, maybe data vis fem?}
Without appropriate context, readers often believe charts present an objective view-from-nowhere~\cite{haraway1988situated}, as opposed to their inherently persuasive and subjective role. A visualization may address the right problem but be doing so using the wrong data set, or there may be a mismatch between assumptions about the data and the data itself, as in \figref{fig:missing-datasets}. The people creating the dataset (or choosing what data are collected) can significantly impact the analytic process~\cite{correll2019ethical,dignazio2019draft,missingdatasets}. 
%
%We acknowledge that there are myriad ways that mirages can arise in the analytics pipeline. Drawing on the taxonomies highlighted here, we compile a table of potential mirages in our supplemental material, sorted by the analytical step we believe is responsible for the resulting failure in the final visualization. We include a subset of this table as Table \ref{table:mirage-table}.
Even representations that present reliable data in a faithful manner are not free from potential mirages. A reader may have previously seen a chart which emphasized facets of the data, anchoring them to their earlier understanding and causing them to misunderstand the current chart. 
There are myriad additional ways that mirages can arise in the VA pipeline. 
In the Table 2 we compile a list of errors that can create mirages, and describe the method by which we compiled those errors. We include a subset in Table~\ref{table:mirage-table}.

\begin{table*}[ht!]
\caption{Examples of errors resulting in mirages along different stages of our analytics pipeline, sorted by the analytical step we believe is responsible for the resulting failure in the final visualization, and colored following \figref{fig:mirage-figure}. This list is not exhaustive, but presents examples of how decision-making at various stages of analysis can damage the credibility or reliability of the messages in charts. A longer version of this table with additional mirages is included in our supplemental materials.}
\ssmall
\begin{tabular}{>{\raggedright\arraybackslash}p{1.8cm}p{15.4cm}}
Error & Mirage\\ \hline
   \rowcolor{colora}\multirow{4}{0em}{\hspace{-0.6cm}\rotatebox{90}{\normalsize{Curating}}}Missing or  Repeated Records  & We often assume that we have one and only one entry for each datum. However, errors in data entry or integration can result in missing or repeated values that may result in inaccurate aggregates or groupings (see \figref{fig:wrangling}). \cite{kim2003taxonomy} \\
 \rowcolor{colora-opaque}Outliers  & Many forms of analysis assume data have similar magnitudes and were generated by similar processes. Outliers, whether in the form of erroneous or unexpectedly extreme values, can greatly impact aggregation and discredit the assumptions behind many statistical tests and summaries. \cite{kim2003taxonomy} \\
 \rowcolor{colora}Spelling Mistakes  & Columns of strings are often interpreted as categorical data for the purposes of aggregation. If interpreted in this way, typos or inconsistent spelling and capitalization can create spurious categories, or remove important data from aggregate queries. (See \figref{fig:misspelling}) \cite{wang2019uni}\\
 \rowcolor{colora-opaque}Drill-down Bias  & We assume that the order in which we investigate our data should not impact our conclusions. However, by filtering on less relevant variables first the impact of later variables can be hidden. This results in insights that address only small parts of the data, when they might be true of the larger whole. \cite{lee2019avoiding}\\

   \rowcolor{colorb}\multirow{4}{0em}{\hspace{-0.6cm}\rotatebox{90}{\normalsize{Wrangling}}}Differing Number  of Records by  Group  & Certain summary statistics, including aggregates, are sensitive to sample size. However, the number of records aggregated into a single mark can very dramatically. This mismatch can mask this sensitivity and problematize per-mark comparisons; when combined with differing levels of aggregation, it can result in counter-intuitive results such as Simpson's Paradox. \cite{guo2017you}\\
 \rowcolor{colorb-opaque}Cherry Picking  & Filtering and subsetting are meant to be tools to remove irrelevant data, or allow the analyst to focus on a particular area of interest. If this filtering is too aggressive or if the analyst focuses on individual examples rather than the general trend, this cherry-picking can promote erroneous conclusions or biased views of the variables. Neglecting the broader data context can result in the Texas Sharpshooter Fallacy or other forms of HARKing~\cite{cockburn2018hark}. \cite{few2019loom}\\
 \rowcolor{colorb}Analyst Degrees of Freedom  & Analysts have a tremendous flexibility in how they analyze data. These ``researcher degrees of freedom''~\cite{gelman2013garden} can create conclusions that are highly idiosyncratic to the choices made by the analyst, or in a malicious sense promote ``p-hacking'' where the analyst searches through the parameter space in order to find the best support for a pre-ordained conclusion. A related issue is the ``multiple comparisons problem'' where the analyst makes \emph{so many} choices that at least one configuration, just by happenstance, is likely to appear significant, even if there is no strong signal in the data. \cite{gelman2013garden,pu2018garden,zgraggen2018investigating}\\
 \rowcolor{colorb-opaque}Confusing Imputation  & There are many strategies for dealing with missing or incomplete data, including the imputation of new values. How values are imputed, and then how these imputed values are visualized in the context of the rest of the data, can impact how the data are perceived, in the worst case creating spurious trends or group differences that are merely artifacts of how missing values are handled prior to visualization. \cite{song2018s}\\

   \rowcolor{colorc}\multirow{4}{0em}{\hspace{-0.6cm}\rotatebox{90}{\normalsize{Visualizing}}}Non-sequitur  Visualizations  & Readers expect graphics that appear to be charts to be a mapping between data and image. Visualizations being used as decoration (in which the marks are not related to data) present non-information that might be mistaken for real information. Even if the data are accurate, additional unjustified annotations could produce misleading impressions, such as decorating uncorrelated data with a spurious line of best fit. \cite{correll2017black}\\
 \rowcolor{colorc-opaque}Overplotting  & We typically expect to be able to clearly identify individual marks, and expect that one visual mark corresponds to a single value or aggregated value. Yet overlapping marks can hide internal structures in the distribution or disguise potential data quality issues, as in \figref{fig:opacity-permute}. \cite{correll2018looks,mayorga2013splatterplots,micallef2017towards}\\
  \rowcolor{colorc}Concealed Uncertainty & Charts that do not indicate that they contain uncertainty risk giving a false impression and may cause mistrust of the data if the reader realizes the information has not been presented clearly. Readers may incorrectly assume that data is high quality or complete, even without evidence of this veracity. \cite{few2019loom, mayrTrust2019, sacha2015role, song2018s}\\
 \rowcolor{colorc-opaque}Manipulation of  Scales  & The axes and scales of a chart are presumed to straightforwardly represent quantitative information. However, manipulation of these scales (for instance, by flipping them from their commonly assumed directions, truncating or expanding them with respect to the range of the data~\cite{cleveland1982variables, correll2019truncating, correll2017black, pandey2015deceptive, ritchie2019lie}, using non-linear transforms, or employing dual axes~\cite{KindlmannAlgebraicVisPedagogyPDV2016, cairo2015graphics}) can cause viewers to misinterpret the data in a chart, for instance by exaggerating correlation~\cite{cleveland1982variables}, exaggerating effect size~\cite{correll2019truncating,pandey2015deceptive}, or misinterpreting the direction of effects~\cite{pandey2015deceptive}. \cite{cairo2015graphics,cleveland1982variables,correll2019truncating,correll2017black,KindlmannAlgebraicVisPedagogyPDV2016,pandey2015deceptive,ritchie2019lie}\\

   \rowcolor{colord}\multirow{4}{0em}{\hspace{-0.6cm}\rotatebox{90}{\normalsize{Reading}}}Base Rate Bias  & Readers assume unexpected values in a visualization are emblematic of reliable differences. However, readers may be unaware of relevant base rates: either the relative likelihood of what is seen as a surprising value or the false discovery rate of the entire analytic process. \cite{correll2016surprise,pu2018garden, zgraggen2018investigating}\\
 \rowcolor{colord-opaque}Inaccessible Charts  & Charts makers often assume that their readers are homogeneous groups. Yet, the way that people read charts is heterogeneous and dependent on perceptual abilities and cognitive backgrounds that can be overlooked by the designer. Insufficient mindfulness of these differences can result in miscommunication. For instance, a viewer with color vision deficiency may interpret two colors as identical when the designer intended them to be separate. \cite{lundgard2019Sociotechnical, plaisant2005information, wutanisszafir2019}\\
 \rowcolor{colord}Anchoring Effect  & Initial framings of information tend to guide subsequent judgements. This can cause readers to place undue rhetorical weight on early observations, which may cause them to undervalue or distrust later observations.  \cite{hullman2011visualization, ritchie2019lie}\\
 \rowcolor{colord-opaque}Biases in  Interpretation  & Each viewer comes to a visualization with their own preconceptions, biases, and epistemic frameworks. If these biases are not carefully considered cognitive biases, such as the backfire effect or confirmation bias, can cause viewers to anchor on only the data (or the reading of the data) that supports their preconceived notions, reject data that does not accord with their views, and generally ignore a more holistic picture of the strength of the evidence. \cite{d2016feminist, dignazio2019draft, few2019loom,valdez2017framework,wall2017warning}\\
\end{tabular}
\label{table:mirage-table}
\end{table*}

\section{Existing Visual Analytics Testing Tools}

Mirages are dangerous because the reader is unaware of them. Automated or semi-automated systems could alleviate this danger by surfacing potential mirages as a way of encouraging data skepticism and re-analysis of the elements underlying a particular chart. An essential focus of our work is developing methods for automatically detecting mirages that occur in the relationship between data and design. In the following section we locate this work within prior techniques for verifying the correctness of analyses at different points in the pipeline.

%In the following sections we explore existing tools for verifying the correctness of analyses and different points in our analytics pipeline. \am{add sentence, something like: though these previous techniques we locate our work}

%There has been previous work on automatically surfacing mirages at some stages in the visual analytic pipeline, such as work from the statistics and database communities on automatically detecting curation  and wrangling  errors \cite{kandel2011research, kim2003taxonomy, stonebraker2013data}.

\subsection{Data Verification}
There are a variety of approaches for automatically detecting data quality issues. Many systems employ combinations of statistical algorithms, visualizations, and manual inspection to detect and correct data quality issues~\cite{kandel2011research}. Most relevant to our approach, Mu\c{s}lu \etal\cite{mucslu2015preventing} employ the metaphor of continuous testing to detect potential data quality concerns, the Vizier system~\cite{brachmann2019data} surfaces data ``caveats'' that might indicate data quality concerns., and Hynes \etal\cite{hynes2017data} propose a data linter and find that many common datasets for use in training and evaluating machine learning models contain elementary data quality issues. Wang \& He~\cite{wang2019uni} propose an automated error detection system for tables based on statistical analyses. Salimi \etal\cite{salimi2018bias} describe a system for automatically detecting bias in analytical queries. Barowy \etal\cite{barowy2018excelint, barowy2014checkcell} present systems for debugging data in spreadsheets. A mixed-initiative data wrangling metaphor is present in a variety of systems~\cite{kandel2012profiler, raman2001potter, stonebraker2013data} as well as in commercial solutions \cite{tableauPrep, trifacta}.

\subsection{Visualization Verification}
Visualization research has not solved the problem of visualization designs responding correctly and clearly to important changes in the underlying data, while not exaggerating trivial changes. While there has been some work from the scientific visualization community on verifying the correctness of images \cite{kirby2008need}, there has been little work~\cite{gotz2019visualization, isenberg2013systematic} on analyzing correctness in basic charts. Rogowitz~\cite{rogowitz2001blair,rogowitz1996not} explore how minor alterations to color maps can result in different perceptions of patterns in visualizations. 
% Also relevant to our work is the use of ``graphical inference''~\cite{wickham2010graphical} where viewers look at multiple visualizations generated under some null hypothesis and one visualization of the actual data, and must identity the ``guilty'' visualization showing the pattern of interest amongst the ``innocent'' null visualizations. 
Wickham \etals ~\cite{wickham2010graphical} ``line-up'' protocol in which viewers look at a collection of charts with randomized data and one with the actual data, and are tasked with identifying the chart containing the real data. 
Hofmann \etal\cite{hofmann2012graphical} use reliability at performing this task as a proxy for the statistical power of a visualization~\cite{hofmann2012graphical}. 
%\am{We haven't set up AVD yet}
%Correll \etal~\cite{correll2018looks} specifically use graphical inference to diagnose algebraic visualization design concerns. 
Visualizations where graphical inference is unreliable suggest that either the statistical pattern of interest is not robust or that the visualization design employed is insensitive to such patterns.
Proposed mixed-initiative solutions to issues of robustness involve supplementing visualizations with additional metrics that indicate their reliability~\cite{binnig2017toward,veras2019discriminability,zhao2017controlling}, or performing pre-analyses to automatically detect potential concerns in a dataset~\cite{guo2017you}. 
%Veras \& Collins~\cite{veras2019discriminability} see robustness as a matter of visualization scalability and deploy a metric-based approach to identify visualization that don't reliably reflect their data . 
%Similarly, Demiralp \etal\cite{demiralp2014learning} use metric-space embeddings of perceptual data as mechanism to automatically identify effective visualizations. 
Lunzer \etal\cite{lunzer2014aint} explore the robustness of a visualization by superimposing alternative chart configurations. 

\subsection{Other Techniques for Visualization Skepticism}
Avoiding known deceptive practices is often instantiated through carefully selected defaults~\cite{few2019loom} in visualization authoring or through recommendation systems (such as in Tableau's ``Show Me''~\cite{mackinlay2007show} or Moritz \etals Draco~\cite{moritz2018formalizing}). To our knowledge, no system exists that automatically detects or surfaces deceptive elements of a visualization design itself.

Even if visualization designs are not deceptive, our cognitive biases can still cause us to make incorrect or unjustified assumptions about the data~\cite{dimara2018task}. Similar to our work, Wall \etal~\cite{wall2017warning} propose a system that automatically augments a visual analytics system with warnings about cognitive biases that may be present in the current course of analysis. In a later work Wall \etal \cite{wall2019toward} describe a design space of strategies for mitigating bias in visual analytics systems. 

%Merely adhering to best practices is not sufficient defense; surfacing requires careful data skepticism and critical thinking. 
%We recognize that 
Mirages can occur in ways that are difficult or perhaps even impossible to detect in an automatic way, relying as they do on potentially idiosyncratic misreadings or omissions in chart interpretation. To that end, automatic methods such as ours could be augmented by tools for introspection that can help identify biases and perceptual problems. D{\"o}rk \etal\cite{dork2013critical} construct a four point system for critically analyzing infographics. Lupi~\cite{lupi2017data} prompts chart makers to reconsider their relationship with their data and rendered image. Wood \etal~\cite{wood2018design} ask visualization designers to engage with potential problems originating at different stages in the design process through linted design-schemas, which ask designers to answer questions from a variety of sources including D'Ignazio \& Klein's~\cite{d2016feminist} \emph{Feminist Data Visualization}, such as \textit{``How do I communicate the limits of my categories in the final representation?" }

\begin{figure}[bth]
   \centering
   \includegraphics[width=0.9\columnwidth]{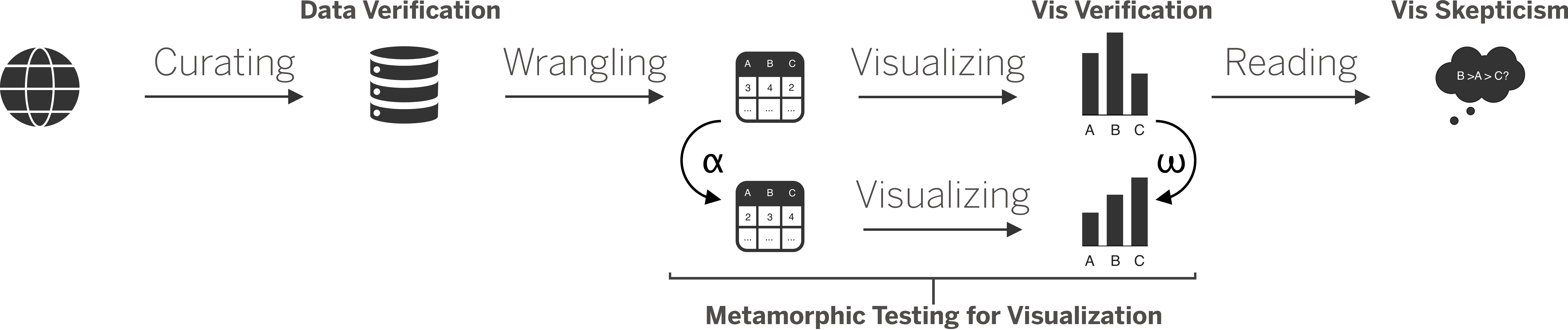}
   \caption{\emph{Metamorphic Testing For Visualization} connects directly to our pipeline in \protect\figref{fig:mirage-figure} by combining algebraic visualization design \protect\cite{kindlmann2014algebraic} and metamorphic testing \protect\cite{segura2016survey}. Our tests alter either the way that the data are manipulated, or the design of the final visualization, with the expectation that (in-)significant changes to the design or data will result in corresponding (in-)significant changes to the final visualization. Failures indicate sensitive or unanticipated choices that can result in mirages.
   }
   \label{fig:mt4v}
\end{figure}

\section{Metamorphic Testing for Visualization}
Our review of the visual analytics testing literature suggests that there has been comparatively less consideration towards detecting errors that occur in the relationship between data and chart, as in \figref{fig:mt4v}. Prior work principally focuses on embedding best practices through automatic chart recommendation rather than validating existing charts. To address this gap we combine a concept from the software engineering community, \emph{metamorphic testing}, which focuses on detecting errors in contexts that lack a truth oracle, with work from Kindlmann \& Scheidegger's Algebraic Visualization Design (AVD)~\cite{kindlmann2014algebraic}, to form a notion of metamorphic testing for visualization.

%\am{Include a comment on why AVD is so good for addressing this type of problem: We focus on AVD because it offers an effective way to organize thinking around particular type of problem}

\subsubsection{Algebraic Visualization Design} 
\begin{figure}[bth]
   \centering
   \includegraphics[width=0.8\columnwidth]{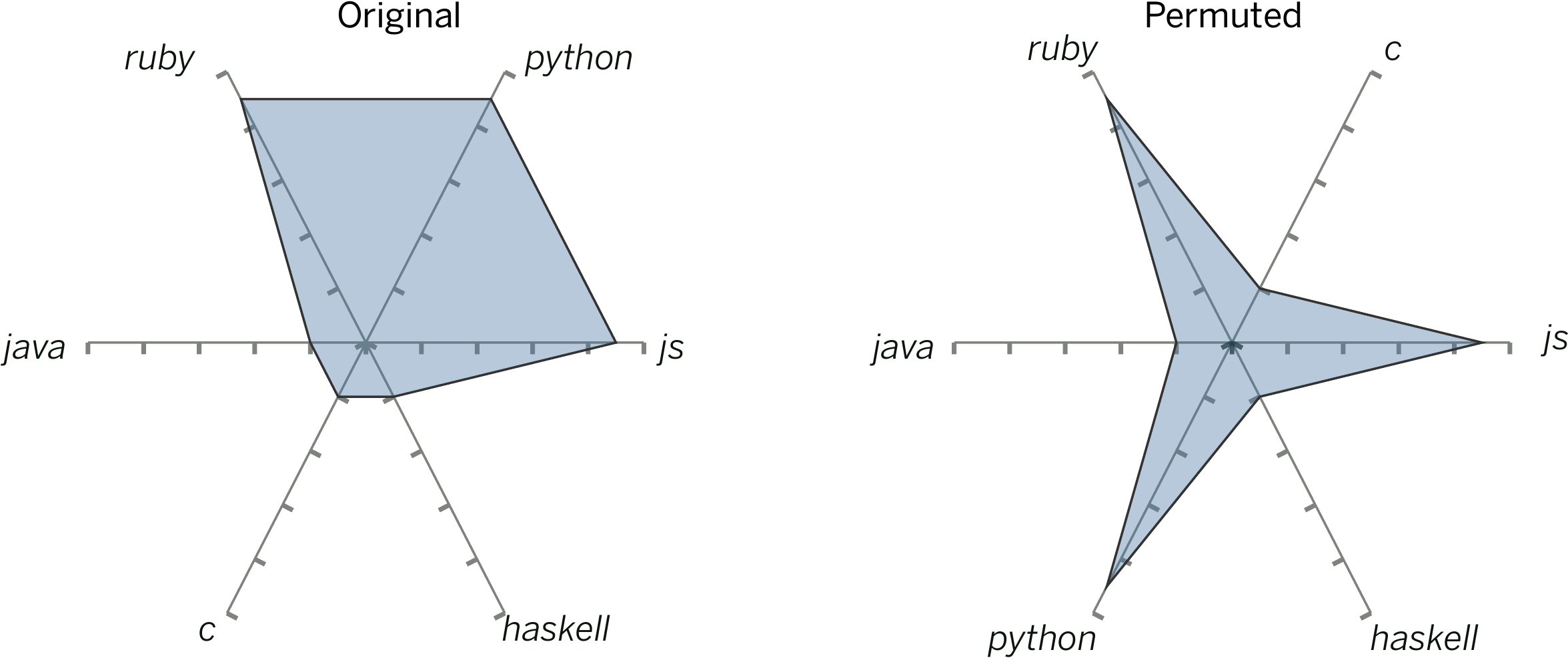}
   \caption{
    Radar charts are often used to compare the skills of job candidates \protect\cite{filipov2019cv3}. Here we show two radars of the programming language skills for a hypothetical job candidate. One axis ordering seems to suggest that the candidate is skilled in one area, while the other suggests that their skills are varied. The change in meaning based on a arbitrary design parameter is an AVD \emph{hallucinator} \protect\cite{kindlmann2014algebraic}.}
   \label{fig:radarsz}
\end{figure}

Under the AVD framework, trivial changes to the data (such as shuffling the row order of input data) should result in trivial changes in the resulting visualization, and important changes in the visual appearance of the visualization should only occur as a result of correspondingly important changes in the backing data. 
These assertions are formalized in a commutativity relation, which describes the properties of an effective visualization across potential data transformations:
%
% TIKZ Commutative diagram from:
% https://tikzcd.yichuanshen.de/#eyJub2RlcyI6W3sicG9zaXRpb24iOlswLDBdLCJ2YWx1ZSI6IkRfMSJ9LHsicG9zaXRpb24iOlsxLDBdLCJ2YWx1ZSI6IlJfMSJ9LHsicG9zaXRpb24iOlswLDFdLCJ2YWx1ZSI6IkRfMiJ9LHsicG9zaXRpb24iOlsxLDFdLCJ2YWx1ZSI6IlJfMiJ9LHsicG9zaXRpb24iOlsyLDBdLCJ2YWx1ZSI6IlZfMSJ9LHsicG9zaXRpb24iOlsyLDFdLCJ2YWx1ZSI6IlZfMiJ9XSwiZWRnZXMiOlt7ImZyb20iOjAsInRvIjoxLCJsYWJlbFBvc2l0aW9uIjoibGVmdCIsInZhbHVlIjoicl8xIn0seyJmcm9tIjoyLCJ0byI6MywidmFsdWUiOiJyXzIifSx7ImZyb20iOjAsInRvIjoyLCJ2YWx1ZSI6IlxcYWxwaGEiLCJsYWJlbFBvc2l0aW9uIjoicmlnaHQifSx7ImZyb20iOjEsInRvIjo0LCJ2YWx1ZSI6InYifSx7ImZyb20iOjMsInRvIjo1LCJ2YWx1ZSI6InYifSx7ImZyb20iOjQsInRvIjo1LCJsYWJlbFBvc2l0aW9uIjoibGVmdCIsInZhbHVlIjoiXFxvbWVnYSJ9XX0=
\begin{equation}\label{equation:commutative}
    v \circ r_2  \circ \alpha = \omega \circ v \circ r_1
    \hspace{0.3 in}
    \begin{tikzcd}
        D_1 \arrow[r, "r_1"] \arrow[d, "\alpha"'] & R_1 \arrow[r, "v"] & V_1 \arrow[d, "\omega"] \\
        D_2 \arrow[r, "r_2"]                      & R_2 \arrow[r, "v"] & V_2                    
    \end{tikzcd}
\end{equation}
Where $D_i$ is the original data, $r_i$ a change in representation, $R_i$ a representation of data, \textit{v} the visualization process, and $V_i$ the resulting image. $\alpha$ is a change to the data which should commute with the corresponding change to the visualization, denoted $\omega$. 
Failures of these assertions can result in ``hallucinators'' (visualizations that look dramatically different despite being backed by similar or identical data, such as in \figref{fig:radarsz}) and ``confusers'' (visualizations that look identical despite being backed by dramatically different data). In the worst case, visualizations can be completely non-responsive to their backing data, functioning as mere number decorations and creating what Correll \& Heer~\cite{correll2017black} refer to as visualization ``non-sequiturs.'' These AVD failures directly tie to our notion of mirages (as they can result in visualizations that are fragile, non-robust, or non-responsive), but, by providing a language of manipulations of data and visualization specification, lend themselves to mixed-initiative or automatic testing. AVD provides a useful framework for designing tests that detect failures that require little domain knowledge. We can simply induce trivial or non-trivial data change, and check for corresponding changes in the resulting visualization.

\subsubsection{Metamorphic Testing}

In complex software systems it can be difficult or prohibitively expensive to verify whether or not the software is producing correct results. In the field of software testing distinguishing between correct and incorrect behaviour is known as the ``test oracle problem''~\cite{barr2014oracle}. The metamorphic testing (MT) ideology attempts to address this challenge by verifying properties of system outputs across input changes~\cite{segura2016survey}. Rather than checking that particular inputs give correct outputs, MT asserts that properties called \textit{metamorphic relations} should remain invariant across all appropriate metamorphoses of a particular data set. MT has been successfully applied to a wide variety of system domains including computer graphics~\cite{metamorphicoopsla17}, deep learning~\cite{segura2016survey}, and self-driving cars~\cite{Zhou2019Metamorphic}. 
%AM: alternate version. MT has been successfully applied to a wide variety of domains~\cite{metamorphicoopsla17, segura2016survey, Zhou2019Metamorphic}. 

We now consider an example from computer graphics for motivation. Donaldson \etal\cite{metamorphicoopsla17} make use of MT to identify bugs in graphics shader compilers. They do so by selecting a shader, making changes to the code that should not affect the rendered image (such as introducing code paths that will never be reached), and checking if the resulting images are the equal after execution.  They formalize this technique by asserting that the following equation should be  invariant:
    \begin{equation}\label{equation:shader}
        \forall x: p(f_I (x)) = f_O (p(x))
    \end{equation}
where $x$ is a given shader program, $p$ a shader compiler, $f_I$  perturbations to the input, and $f_O$ changes to the output (usually the identity under their framework). The definition of equality in MT plays a significant role in the effectiveness of its analysis. Donaldson \etal use $\chi^2$ distance between image-histograms as a proxy for image equality. Using this approach they found over 60 bugs in commercial GPU systems.

\subsection{Applying Metamorphic Testing}

\begin{figure}[t]
   \centering
   \includegraphics[width=0.9\columnwidth]{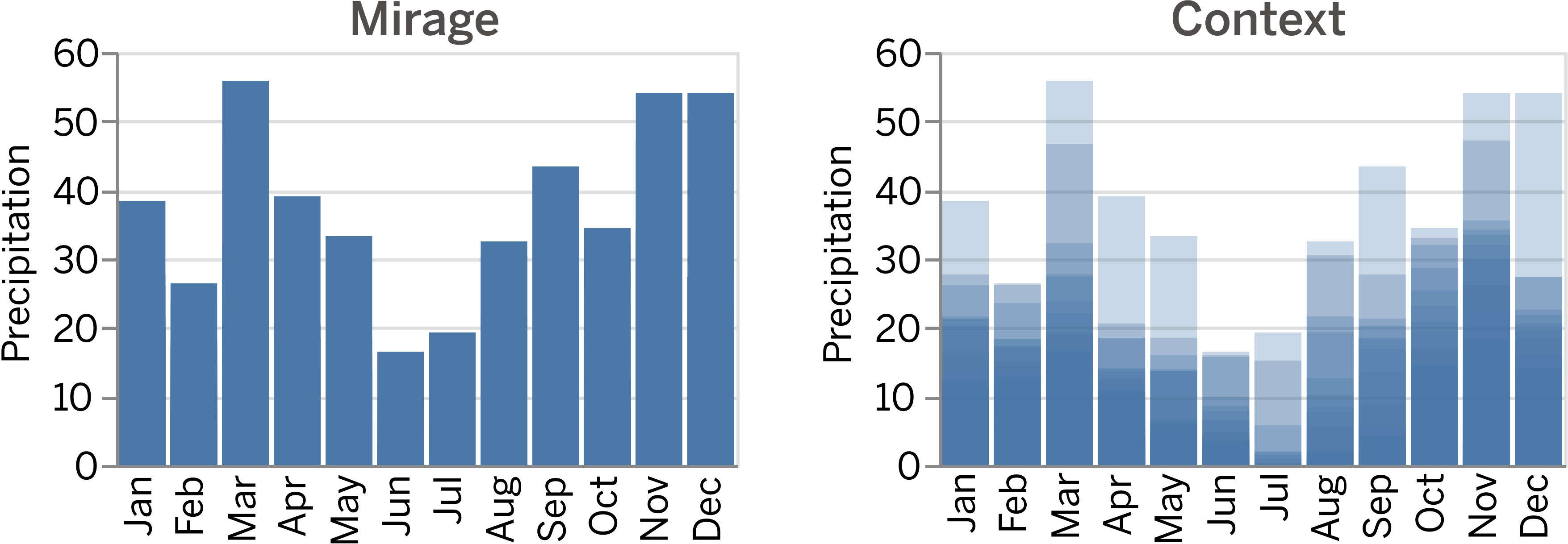}
   \caption{
The lack of a set aggregation for the precipitation axis results in many bars being overplotted in the same visual space (and so causing a potential misinterpretation as only the maximum value for each category is visible). While we could test for this error by directly consulting the visual specification of the chart, not all charts (or even bar charts) are inherently invalid simply because they fail to include an aggregate. A morphism could detect this issue without this constraint. For instance, we would assume that reducing the opacity of marks in a chart would result in a visually similar chart to the full opacity version. A violation of this assumption indicates overplotting.
   }
   \label{fig:opacity-permute}
\end{figure}

We now introduce the idea of use metamorphic testing as a mechanism to verify individual visualizations. Tang \etal ~\cite{tang2019towards} describe visualization as the function  $vis(Data, Spec)$. 
% Alternatively: VisNet etal describe visualization as function consisting of triples of task/specification/data.
This suggests two key aspects across which we can execute metamorphic manipulations: alterations to the data and alterations to the design specification. This perspective has the advantage that we can test a wide variety of types of visualization without knowing much about the chart being rendered. For instance, in \figref{fig:opacity-permute}, introducing a trivial \emph{morphism} (in this case a reduction in mark opacity) with the expectation that it should have relatively little change on the resulting graph reveals a chart error.
We observe that \eqnref{equation:shader} is isomorphic to AVD's commutativity relation, \eqnref{equation:commutative}.
%~\cite{kindlmann2014algebraic}, which describes properties of an effective visualization across potential data transformations:
MT is a concrete way to test the invariants of systems in general, whereas AVD describes the types of invariance-failures that occur with visualizations specifically.
%This suggests a link between AVD and MT, in that metamorphic testing appears to be a concrete way to test invariance of systems in general, whereas AVD enumerates theoretical framework that specifies types of invariance-failures that occur to visualizations specifically. 
Observing this overlap we define a \textbf{Metamorphic Test for Visualization (MTV)} as a function parameterized by an equality measure ($\textit{Eq}$), an input perturbation ($\alpha$), a visual perturbation ($\omega$), which evaluates a tuple of data and chart specification (denoted as a pair as $x$), and returns a Boolean. We describe this function in pseudo-Haskell:
\begin{equation}\label{eqn:mtv}
    \begin{aligned}
    MTV&:: (\textit{Eq}, \alpha, \omega) \Rightarrow (spec, data) \Rightarrow Boolean\\
    MTV &(\textit{Eq}, \alpha, \omega) x = Eq(v(\alpha(x)), \omega(v(x)))
    \end{aligned}
\end{equation}
We leave $v$, the visualization system, out of the parameterization because we are interested in testing for problems in the relationship between data and chart specification, as opposed to validating the system mapping chart specification to data space (which we assume to be error free). 
This formulation clearly describes the relationship between expectation and permutation in a manner that we believe allows for concise and unambiguous descriptions of invariance tests.
%This definition allows us to express a wider array of types of test in a less ambiguous manner. 
%This formulation of metamorphic testing also bares a resemblance to Demiralp \etals work on perceptual kernels, which look for small changes to data to be reflected in small changes to perceptual effect, and large data changes to be reflect in large visual changes \cite{demiralp2014learning}. 
%Veras and Collins unify these two approaches in the context of complexity metric driven technique for analyzing the scalablity of a visualization \cite{veras2019discriminability}.

To our knowledge MT has not previously been used in visualization contexts, though there has been prior work that uses implicitly related techniques.
Guo \etal\cite{guo2017you} use a metamorphic-like strategy to detect instances of Simpsons's paradox in a visual analytics system. 
McNutt \etals~\cite{mcnuttlinting} visualization linting system touches on MT-adjacent techniques as a way to identify some AVD failures.
Chiw \etal\cite{chiw2017datm} use MT to validate the correctness of a compiler for a scientific visualization DSL.
Our approach is closely related to techniques that use bootstrapping, randomization, or other statistical procedures to reveal various properties \cite{anand2015automatic, barowy2018excelint, barowy2014checkcell, matejka2017same}, such as Gotz \etals~\cite{gotz2019visualization} ``Inline Replication'' analysis of the visual impact of ``alternative'' analyses and tests for the reliability of a given chart, or Dragicevic \etals\cite{dragicevic2019increasing} ``Multiverse Analysis.''

%To our knowledge MT has not previously been used explicitly in the context of data visualization, though there have been a variety of works related to visualization that touch on explicitly or use implicitly related techniques.  
%Ramanathan \etal make use of metamorphic technique in conjunction with visualization in order verify implementations of epidemiological models \cite{ramanathan2012verification}. 
%Our approach resembles Matejka \etal~\cite{matejka2017same} stat-preserving data perturbation techniques  as well Anand \etals~\cite{anand2015automatic} randomization techniques, although we deploy them to inform binary testing mechanisms. Also 
% Our approach resembles Matejka \etal~\cite{matejka2017same} stat-preserving data perturbation techniques  as well Anand \etals~\cite{anand2015automatic} randomization techniques, although we deploy them to inform binary testing mechanisms. Also related are various techniques that use bootstrapping or other morphisms to reveal the visual impact of ``alternative'' analyzes and test for the reliability of a particular visualization, such as Gotz \etals~\cite{gotz2019visualization} ``Inline Replication'', Dragicevic \etals~\cite{dragicevic2019increasing} ``Multiverse Analyses'', and Barowy \etals~\cite{barowy2014checkcell} approach to debugging spreadsheets.

\begin{figure}[t]
   \centering
   \includegraphics[width=0.9\columnwidth]{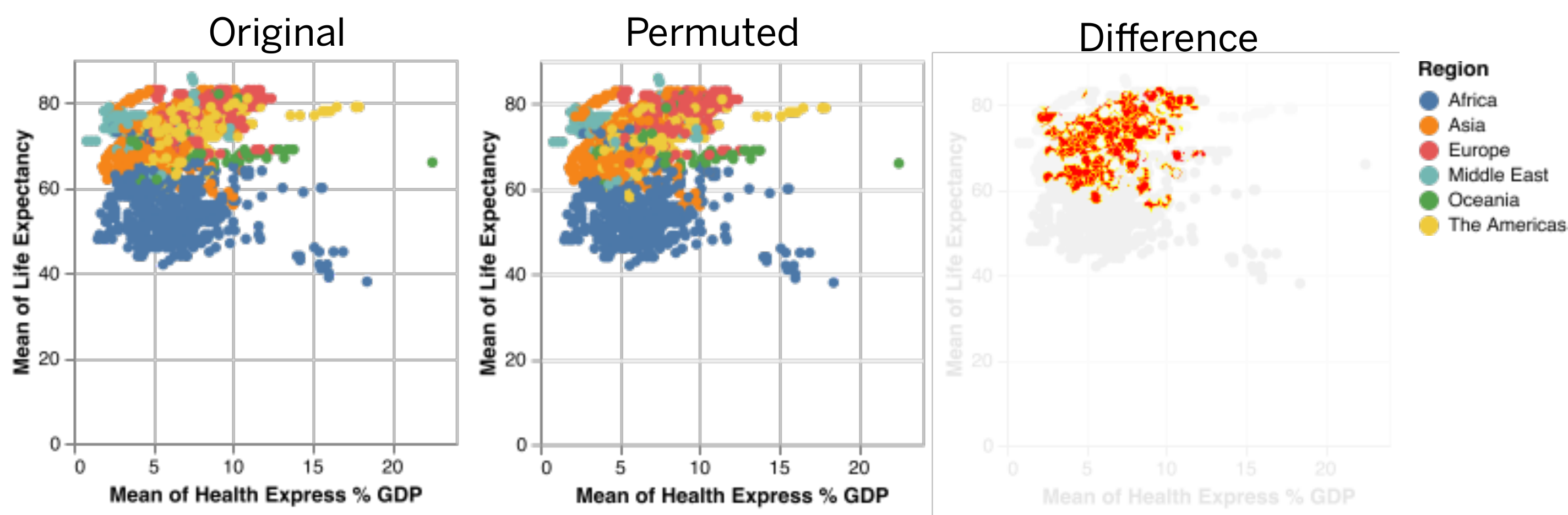}
   \caption{Shuffling the input data can reveal potential mirages. Here we consider a scatterplot drawn from the ``World Indicators" dataset~\protect\cite{worldBank} (left), we permute the input data (center), and construct a the pixel difference (right). This reveals a property of this spec and data combination that it is not resilient to order permutation. In the language of AVD, this chart has a hallucinator. The difference here is due to overdraw: interior regions of the central cluster may or may not be visible among the dominant classes depending on the order in which the data are rendered.
   }
   \label{fig:shuffle-lint}
\end{figure}

\subsection{Proof Of Concept}
We implemented a proof of concept system for inducing morphisms on static Vega-Lite~\cite{satyanarayan2016vega} specs and their backing data in order to identify potential mirages or unreliable signals in charts. Our primary goal in this system is to demonstrate the validity of our metamorphic testing concept. 
%AM: unclear if necessary. We implement a small number of metamorphic relations that validate charts drawn from a static subset of vega-lite and show examples of our system working over synthetic and real-world datasets. 
Our proof of concept focuses on Vega-Lite because of its advantageous API, although our techniques are applicable in principle to any charting system. 
In the following subsections we present a set of metamorphic tests for visualization (MTVs). Each test should have predictable impacts on the resulting image. 
Failing to adhere to a prediction (and hence violate an MT relation) can indicate an error in the backing data or visual specification of the chart, pointing to a potential mirage.
%Failing to adhere to a prediction (and hence violate an MT relation) can indicate an error in the backing data or visual specification of the chart, pointing to a potential visualization mirage.
We include a visual explanation of each of the transformations involved in the following tests in \figref{fig:transformation-explainer}.
  
\subsubsection{MTV: Shuffle} We assert that changes to the order of the input data should not change the rendered image. 
Our detection technique is a pixel differencing algorithm for which we select a tunable threshold in order to reduce the number of false-positives.
More formally, in this test we take $\textit{Eq}$ to be a maximum number of pixels differing between the rendered images, $\alpha$ to be a permutation of the order of the input rows, and $\omega$ to be the identity. 
This test allows us to detect over plotting, as exemplified in \figref{fig:shuffle-lint}, as drawing overplotted groups in different orders will result in visually different charts. Not all overplotting is necessarily indicative of a mirage, but alerting the user to its presence can be useful across many chart types.

%\subsubsection{MTV: Randomly Delete} 
%\am{TODO: replace with contract}
%Mackinlay~\cite{mackinlay1986automating} asserts that for a visualization to be effective it should encode all the facts in the dataset, and encodes only those facts . Here we tests this property by taking our $\alpha$ to be removing a randomly selected third of the rows in the input table, $\omega$ to be the identity, and $\textit{Eq}$ to be a greater than $\epsilon$ pixel difference. That is, if we remove a substantial amount of the data in the chart, the chart should change. Through this metamorphism we are able to detect AVD confusers, such as Correll \etals~\cite{correll2017black} non-sequitur visualizations, and visualization with inappropriately extreme aggregates and some sorts of overdraw, such as the one described in \figref{fig:opacity-permute}.

\begin{figure}[t]
   \centering
   \includegraphics[width=\columnwidth]{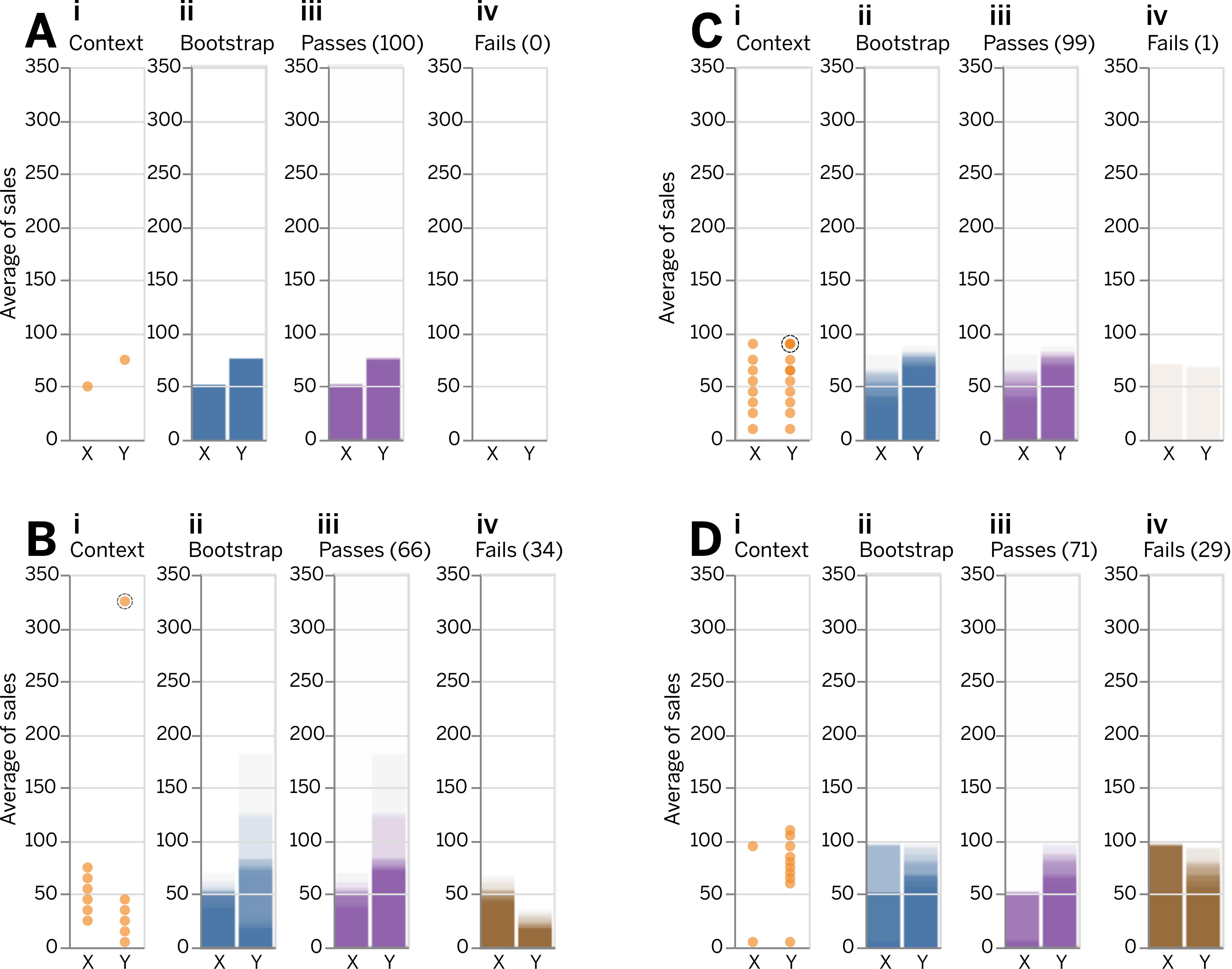}
   \caption{
    We apply our \textit{MTV: Bootstrap} test to to the quartet of distributions shown in \figref{fig:deception-quartet}, shown here in (i). In each case we execute our test $N=100$ times. We create 100 new potential bar charts by resampling from all the data that are aggregated into each bar. We show the results of this test in (ii-iV) by overlaying each low opacity ``potential'' bar chart on top of each other. Large ``fuzzy'' bands indicate that the specific values in the chart are not robust to resampling. We sort our bar charts into two categories: ``passes'' (iii) where the original insight that AVG X > Y is preserved, and ``fails'' where this insight is not preserved (iv). A high number of passes indicate a robust insight.
    }
   \label{fig:deception-quartet-mt}
\end{figure}

\subsubsection{MTV: Bootstrap} 
We assert that the apparent patterns in visualizations should be robust: that is, a particular relationship should continue to be present across minor changes~\cite{correll2018looks,lunzer2014aint}. In this test we focus on bar charts as it allows us a greater degree of nuance in constructing our detector. We take $\textit{Eq}$ to be the same order of heights in the bar chart, $\alpha$ to be a bootstrap sample~\cite{efron1992bootstrap} of all the rows within each of the bars in the chart, and $\omega$ to be the identity.  Bootstrapping allows us to test for variability in a relatively parameter-free way across a wide variety of data distributions and complexities.
%
%We identify which input rows to modify by creating a notion of backward provenance~\cite{wu2013scorpion} that links each aggregate mark to the input tuples that describe it.
We identify which input rows to modify through a backward provenance algorithm~\cite{wu2013scorpion} that links each mark to the input tuples that describe it.
Because bootstrapping relies on random sampling, we adapt our metamorphic testing to statistical view, in which we execute \eqnref{eqn:mtv} a large $N$ number of times and define a pass as a sufficiently large $\epsilon$ fraction of passing sub-tests.
To our knowledge this approach of using aggregated randomized metamorphisms is novel within metamorphic testing, though it bares a close resemblance to Guderlei \etals\cite{guderlei2007statistical} statistical metamorphic testing, which tests functions containing randomness as opposed to using randomness to test functions as we do.
This application of the bootstrap to visualization validation also bares a close resemblance to Gotz \etals\cite{gotz2019visualization} Inline Replication technique, but focuses less on the variability of a particular measure but more on the fragility of the actual visualization itself.
Through this technique we are able to identify when visualizations are liable to be dependent on outliers or small number of divergent records are driving differences between aggregates, as in \figref{fig:deception-quartet-mt}.
The specific tuning of $N$ and $\epsilon$ is task, application, and encoding dependent and warrants further investigation.

\subsubsection{MTV: Contract Records}
\figref{fig:deception-quartet} demonstrates how aggregates can usefully summarize information but they can also mask data problems, such as differing number of records, sampling issues, and repeated records. In this test we examine the robustness of measures in the context of potentially dirty data. Just as in the previous test, we focus on categorical bar charts and take $\omega$ to be the identity, and  $\textit{Eq}$ to be bar height order. Our new $\alpha$ identifies the minimum number of records that make up a bar, and contracts the number of records constituting all other marks down to that minimum through sampling without replacement. Just as in the previous test we also deploy a randomization procedure to probe the central tendency of this measure. If all bars have similar samples sizes, and this sample size is sufficiently large, and the aggregation method sufficiently robust to extreme values, this procedure ought to result in reasonably similar charts. This test therefore allows us to detect variability caused by sampling issues and other problems relating to differing number of records. Additionally, through this morphism we are able to detect some additional AVD confusers, such as Correll \etals\cite{correll2017black} non-sequitur visualizations and some sorts of overdraw, such as the one described in \figref{fig:opacity-permute}, as non-responsiveness to removing substantial amounts of data indicates a chart's insensitivity to its backing data.

\subsubsection{MTV: Randomize}
Parameterized tests may not capture subtle relationships between variables. Anand \etal\cite{anand2015automatic} use randomized non-parametric permutation tests to assess the relative likelihood of different visual patterns in scatterplots. We adopt this test by randomizing the relationship between two variables. As with the prior two tests, we focus on categorical bar charts, taking $\omega$ to be the identity, and  $\textit{Eq}$ to be bar height order. Our $\alpha$ is then a random permutation of the value and category assignments. Unlike with the previous tests, we expect that if the signal is not particularly robust, the charts will be relatively \textit{similar}: destroying the relationship between variables would not change the chart much. A high proportion of highly dissimilar charts indicates significant relationships between category and value. This test can reveal mirages related to sampling error and signal-to-noise ratios. The test described in Figure 11 might also be achieved through randomization.

\subsection{Simulation Results}
We conducted a set of simulations to see whether mirages caused by data errors could be reliably detected by our metamorphic tests. To that end, we generated a series of synthetic data sets. Similar to Zgraggen \etals\cite{zgraggen2018investigating} test on the reliability of insights from visual analytics, each data set consisted of two subsets sampled from two Gaussian distributions, $X$ and $Y$, with $n=50$, $\mu=50$, $\sigma=10$. These points were visualized as a categorical bar chart of means, as in \figref{fig:deception-quartet}. We would expect any difference in the height of the bars to be non-robust and unreliable; any significant differences between the two categories would be a mirage. To assess the utility of our metamorphic tests we then varied the parameters of the generating Gaussian for $Y$ to induce more or less robust group differences.

\begin{enumerate}[(i)]
    \itemsep0em
    \item \emph{mean}: We vary $\mu$ for $Y$'s Gaussian. We expect this to produce more ``reliable'' differences and that \emph{MTV: Randomize} will excel at identifying this change.
    %\item \emph{sample size}: Instead of 50, only (25,20,15,10,5) points were generated for one category. We expected {MTV:Contract Records} to excel at identifying this large differences in sample size. 
    \item \emph{sample size}: We vary $n$ of $Y$'s Gaussian. We expect \emph{MTV:Contract Records} to excel at identifying large differences in sample size. 
    \item \emph{outlying values}: We add $k$ outliers to $Y$, generated by sampling uniformly from $[1.5\times IQR+Q_3,3\times IQR+Q_3]$. We expect \emph{MTV: Bootstrap} to excel at this task.
    \item \emph{variance}: We vary $\sigma$ of $Y$'s Gaussian. We expect \emph{MTV: Bootstrap} to excel at identifying the increased variability.
\end{enumerate}

%AM: trying out a version without numbers?
%\begin{description}
%    \itemsep0em
%    \item[\emph{mean}]We vary $\mu$ for $Y$'s Gaussian. We expect this to produce more ``reliable'' differences and that \emph{MTV: Randomize} will excel at identifying this change.
    %\item \emph{sample size}: Instead of 50, only (25,20,15,10,5) points were generated for one category. We expected {MTV:Contract Records} to excel at identifying this large differences in sample size. 
%    \item[\emph{sample size}]We vary $n$ of $Y$'s Gaussian. We expect \emph{MTV:Contract Records} to excel at identifying large differences in sample size. 
%    \item[\emph{outlying values}]We add $k$ outliers to $Y$, generated by sampling uniformly from $[1.5\times IQR+Q_3,3\times IQR+Q_3]$. We expect \emph{MTV: Bootstrap} to excel at this task.
%    \item[\emph{variance}]We vary $\sigma$ of $Y$'s Gaussian. We expect \emph{MTV: Bootstrap} to excel at identifying the increased variability.
%\end{description}

%We generated 30 datasets for each set of alterations, for a total of $4 \times 3 \times 30 = 600$ datasets. We then tested each resulting visualization using \emph{MTV: Contract Records}, \emph{MTV: Randomize}, and \emph{MTV: Bootstrap}. We excluded \emph{MTV: Shuffle} as the row-order of values in an aggregated categorical bar charts did not produce any variability in output.
We generated 30 datasets for each of the 4 alterations across 5 effect sizes, for a total of $30 \times 4 \times 5 = 600$ datasets. 
We then tested the resulting charts with each of the MTVs described above, excluding \emph{MTV: Shuffle}, which did not yield any variability in the output. Our datasets, as well as a prototype tool for exploring our tests, are available at \url{https://osf.io/je3x9}.

\begin{figure}[t]
   \centering
   \includegraphics[width=\columnwidth]{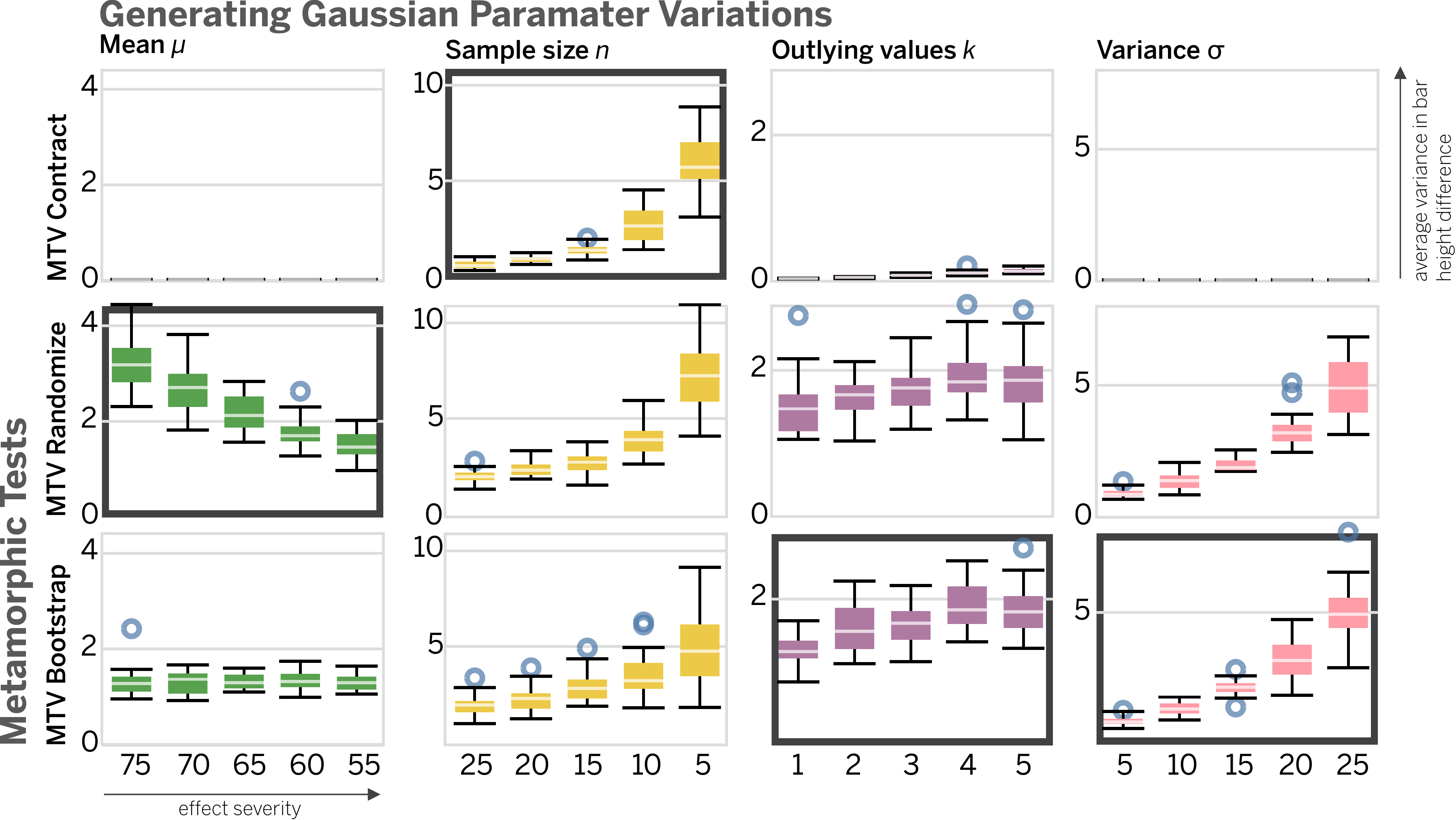}
   \caption{
    %The results of our metamorphic tests on 600 simulated two-column bar charts. We ran each test following our randomized MT approach with $N=100$, and measured the variance in difference in bar height. Each column is a different parameter we varied when generating the data, and each row is a different metamorphic test. Boxes around the cells indicate tests we expected to be especially relevant for detecting the relevant manipulation. The y-axis for each chart is the variance in bar height. High variance indicates that the bar values are highly unstable or unreliable. A metamorphic visualization linter would flag these visualizations for further examination.
    The results of our metamorphic tests on 600 simulated two-column bar charts (bar charts not-shown). We ran each test following our randomized MT approach with $N=100$, and measured the variance in difference in bar height. A metamorphic visualization linter would flag charts  with high variance flag for further examination. Note that in the sample size variation higher effect size means a lower number of points, this emulates under-sampled data. 
   }
   \label{fig:eval-tests}
\end{figure}

\figref{fig:eval-tests} shows the results of our simulation. Each column is a different parameter we varied when generating the data, and each row is a different MTV. Boxes around the cells indicate tests we expected to be especially relevant for detecting the relevant manipulation. The y-axis for each chart is the variance in bar height. High variance indicates that the bar values are highly unstable or unreliable. 

In general, the impact of our morphisms became larger as the severity of our data manipulations increased: the fragility of the values in a given bar chart increases as the means become closer together, the sample size shrinks, outliers are added, or the variability increases. The exception is the randomize test, where we would expect less variability as the two distributions become more similar: high variability in this case is an indication that there is a true signal that is being disrupted by our morphism. As with AVD, we expect significant changes to our data to result in correspondingly significant changes in our charts: failures to do so should invite skepticism in the viewer.

While we recognize that our simulation does not fully capture the utility of our proposed metamorphic tests, we present these initial results as evidence that our tests can be used as measures for the robustness of signals in visualizations.

\section{Discussion}

We believe that MT offers a useful complement to directly testing data or chart specifications, as it 
%is both optimistically na\"ive, 
requires a smaller set of assumptions and parameters than statistical tests, and is portable across visualization toolkits.
We see the types of visualization tests described here as being analogous to testing methods from software engineering. Direct tests, like unit tests, verify isolated properties (for instance, that quantitative axes begin at `0' in bar charts); while metamorphic tests, like integration tests, look to see that the whole image is working as desired. We believe that, in tandem, these validation approaches offer an effective way to target a wide variety of charting errors arising in the Wrangling and Visualizing steps of visual analytics.
%
%A critical assumption in MTV is that the visualization under consideration is working as desired. This assumption allows us to investigate things we couldn't have assumed were wrong: direct tests can be written for problems that are known to exist, but the permutative nature of MT allows us access to a interesting class of potential errors. For instance, bootstrapping the entire input data in \figref{fig:opacity-permute} reveals that the data wasn't aggregated. While this is easily tested for directly, it is hard to know that this could have been an error to begin with (and in fact is motivated by an error one of those authors made while creating that very chart).
% \am{Add to discussion: this work is the first foray into an interesting space of basic charting validation. There are a number of open challenges to expand upon from here. Big door}
This work is a first foray into an investigation into mixed-initiative verification of visualizations. There are a number of interesting challenges in this space including effective presentation of automated results, development of faster and more effective analysis techniques, and capturing additional domains and tasks. Even so, our testing regime can be extended to new instances of known visualization biases: there are many morphisms we can induce to test for areas of concern. For instance, many choropleth maps are not particularly informative~\cite{correll2016surprise}: by replacing the data with base rates, we can test for the strength of geospatial trends. As with graphical inference~\cite{hofmann2012graphical,wickham2010graphical}, by replacing the data in a given chart with data generated under a null hypothesis, we can test for the detectability of important patterns.
%
%To this last point we observe that, beyond the small number of examples that we described in the preceding section, our framework can be extended to new instances. For instance, Correll \& Heer \cite{correll2016surprise} describe how choropleth maps are often not normalized. This can result in showing only the base rate or high variability due to low-sample size. Both errors fall within the regime of MTV, and could be tested for swapping out the initial data with an equivalent base rate data (such as the underlying population) and comparing the order of the states associated with the color measure. 

\subsection{UX/UI Challenges}
Software analysis systems are only effective if they catch errors in a manner that improves the quality of the work being performed, which is contingent on being trusted by their users.
Our proof-of-concept system follows the interface pattern of a software linter. Linters are a type of software analysis tool that usually employ static analysis to catch semantic and stylistic programming bugs, like a spell-checker for code~\cite{johnson1977lint} (although some lint systems have moved into non-programming domains ~\cite{barowy2018excelint,writegood, hynes2017data, proselint, mcnuttlinting, qu2017keeping, wood2018design}).
Like McNutt \& Kindlmann~\cite{mcnuttlinting}, we find linters to be a useful paradigm for describing correctness in visual analytics. They are typically designed with the perspective that it is better for the user to be alerted to a non-existent bug (false-positives) than to miss a real one (false-negative), and allow the user to opt out of particular checks when they know better.
Jannah~\cite{jannahmetareader} explores a linting metaphor for alerting users to data quality issues preceding data exploration.
We believe that this type of granular and polite~\cite{whitworth2005polite} control over analysis is a good fit for the level of detail and accuracy that our system can provide.

The optimal UI paradigm for expressing these computationally measured notions of correctness requires future research. While we believe that linters are a strong first foray into this topic, they are not without flaws.
Srinivasan \etal\cite{srinivasan2018augmenting} construct a system which presents statistical facts relevant to individual charts across the data exploration process. Users tend to interpret the presence or absence of these facts as endorsement or criticism. 
Future systems will need to carefully mitigate false-positives (so users do not ignore advice when it is valuable) and to clearly articulate false-negatives (so users know when to act on system output).
As Sacha \etal\cite{sacha2015role} point out, striking the right balance is critical for maintaining user trust in the system.
%
%An important challenge is designing analytic rules for detecting those problems that meaningfully change the message of the chart (mirages), and verifying that those rules work in practice to reduce mirages.
An important challenge is designing analytic rules that detect problems that meaningfully change the message of the visualization (mirages), and verifying that those rules work in practice. 
This is in contrast to rules derived from aesthetic preference, which are common in collections of guidelines, such as rules disallowing pie charts~\cite{diehl2018visguides, mcnuttlinting, meeksplotcon} regardless of their effectiveness for some tasks~\cite{redmond2019visual}.

\subsection{Limitations \& Future Work}

An appealing component of lint systems is that they are typically very fast.
Our current methodology relies on bootstrapping and other statistical techniques which can cause a significant delay in the user receiving feedback (sometimes up to tens of seconds for very large data). Constructing a visualization linting system that addresses these performance challenges (perhaps in the vein of {Mu{\c{s}}lu} \etals \cite{mucslu2015preventing} continuous data integration system) is an intriguing systems problem.
Some types of mirages do not make sense to metaphorically test. For instance, Pandey \etal\cite{pandey2015deceptive} describe that flipped axes can lead to flipped understandings of the real message. While it is possible to design a metamorphic test to identify this type of mirage, it is simpler to query the chart specification directly, rather than induce a morphism and test for difference. 
Some of our tests address errors that are already well known and well studied, such as overplotting~\cite{micallef2017towards}. Some of our tests involve image diffing or other burdensome computations, which will likely be slower and more prone to error than an equivalent system for directly testing for overplotting.

In future work we would like to more fully develop our tool to validate a wider range of chart designs and types. We believe it would be most useful to apply our system to ad hoc charting systems, such as Altair \cite{vanderplas2018altair} or LitVis \cite{wood2018design}, which both consume vega-lite as charting engine. Following Donaldson \etal \cite{metamorphicoopsla17}, our examples focused on tests where $\omega$ is set to be to be the identity for simplicity. In future work we intend to explore the class of $\alpha$s that have predictable and computationally measurable $\omega$s that are not equal to the identity. Visualization linters could be deployed as a continuous integration step that would verify that publicly displayed charts are mirage free.%, in the vein of {Mu{\c{s}}lu} \etals \cite{mucslu2015preventing} notion of continuous data integration.

%Ideally, linters similar to our prototype could be deployed in a continuous integration step that would check that charts being displayed in public venues are mirage free, in the vein of {Mu{\c{s}}lu} \etals \cite{mucslu2015preventing} notion of continuous data integration.

The full space of visualization mirages is vast, and covers complex ground like critical reasoning, cognitive biases, and inequality. There are some mirages that may never be amenable to testing or verification, especially not in as straightforward a way as issues driven by outliers or sampling error. Even for the subset of mirages for which testing is appropriate, extending MT to other parts of our pipeline model may require new assessment techniques. For instance, Kong \etal\cite{kong2018frames} explore how differing titles affect comprehension of data, and Xiong \etal\cite{xiong2019curse} explore how different primings about the data domain can bias how the data are interpreted. These sorts of morphisms can directly influence the creation of mirages but may be hard to algorithmically detect. We suggest handling this with a mixed initiative process of visualization certification, in which users answer questions about visualizations that have had  various morphisms automatically applied to their data or chart specifications.
%\am{@MC i don't get this cite, i support doing it, but i don't really understand why its here}
%Similarly, for cases where verification or correctness is tied to specific statistical regimes, we might make use of specific statistics DSLs, as in Jun \etal~\cite{jun2019tea}.

\subsection{Conclusion}

In this paper we introduce the idea of a \textit{visualization mirage}: a visualization that provides an inference which, upon more detailed examination, disappears or is cast into doubt. To understand the origin of mirages we construct a conceptual model for identifying causal links between choices made in the visual analytics process and the downstream effects on reader comprehension.
We improve on prior work on deceptive visualizations by describing errors that propagate across the visual analytic process and that are not encapsulated in a single aspect or part.
Through this collection of ideas we describe a landscape of issues in the visual analytics process, the problems to user understandings they cause, and how they might be resolved.
To address this final point, we introduce the idea of using \textit{Metamorphic Testing} as a mechanism for automatically detecting mirages arising from the relationship between data and visual encoding.
We provide evidence of the validity of this idea by constructing a prototype system that is able to discern an intriguing class of errors. 
We believe our model and testing approach provide ample starting ground for future work on automated detection of subtle errors in visualization, as well as validating the design of visualizations based on the relationship between their data and design.

\section{Acknowledgments}

%Acknowledgments omitted for anonymous review.
%We thank our anonymous reviewers for their thoughtful commentary, as well as Muareen Stone and Madeleine Thompson for their helpful discussions.
We thank our anonymous reviewers,  as well as Muareen Stone and Madeleine Thompson, for their thoughtful commentary.
% AM: Who else do we thank?????? Maureen? Some other internship-based paper i've seen include a line like "this work performed during an internship" do we include that type of thing here?

% Balancing columns in a ref list is a bit of a pain because you
% either use a hack like flushend or balance, or manually insert
% a column break.  http://www.tex.ac.uk/cgi-bin/texfaq2html?label=balance
% multicols doesn't work because we're already in two-column mode,
% and flushend isn't awesome, so I choose balance.  See this
% for more info: http://cs.brown.edu/system/software/latex/doc/balance.pdf
%
% Note that in a perfect world balance wants to be in the first
% column of the last page.
%
% If balance doesn't work for you, you can remove that and
%  hard-code a column break into the bbl file right before you
% submit:
%
% http://stackoverflow.com/questions/2149854/how-to-manually-equalize-columns-
% in-an-ieee-paper-if-using-bibtex
%
% Or, just remove \balance and give up on balancing the last page.
%

% BALANCE COLUMNS
\balance{}

% REFERENCES FORMAT
% References must be the same font size as other body text.
\bibliographystyle{SIGCHI-Reference-Format}
\bibliography{proceedings}

\appendix

\begin{figure*}[h!]
   \centering
   \includegraphics[width=2\columnwidth]{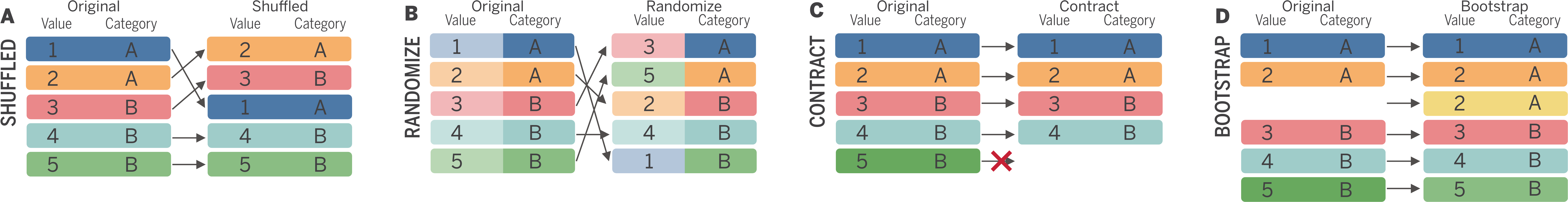}
   \caption{
    A visual explanation of the morphisms used in each of the metamorphic tests described in the main paper. In each section of the figure we start off with an initial table of values, marked Original, and then perform an example transformation, marked accordingly. As each of these transformation involves randomness, we can only make test-level assertions against them in the aggregate. 
    }
   \label{fig:transformation-explainer}
\end{figure*}

\begin{figure}[h]
   \centering
   \includegraphics[width=\columnwidth]{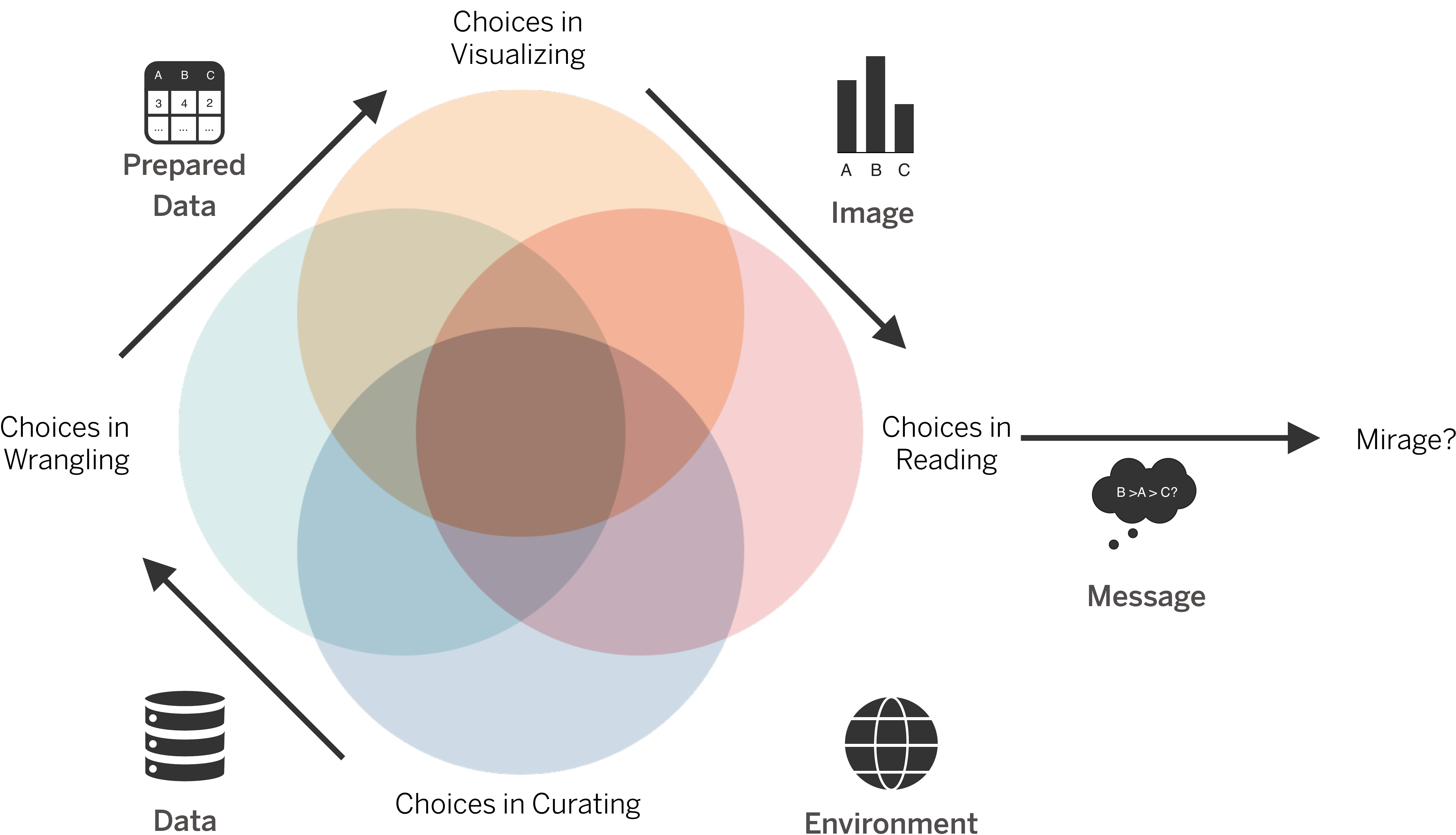}
   \caption{
    A variation on the visual analytics pipeline discussed in the main paper. In this version of the diagram we emphasize the way that errors in different stages can coalesce and interact to provide various sorts of errors.
   }
   \label{fig:supplement-view}
\end{figure}

This appendix includes an expanded version of Table \ref{table:mirage-table} from the main paper, shown here as Table 2. For reasons of space we could include only a small set of visualization mirages in the main paper: we expand on that list here, drawing from potential errors that can occur in more steps of the visual analytics pipeline, as in \figref{fig:supplement-view}. While this expanded table includes additional examples, we recognize that many components of these mirages draw on entire fields of inquiry from statistics, cognitive psychology, and critical theory. As such, we do not claim that this table is complete either, but simply a more exhaustive list of errors, guided by existing work in visual analytics research.
The categories in this table and the papers that constitute it were assembled through an iterative and organic search process.  We created a series of successive models which captured different aspects of the types of errors described in various papers, but settled on the pipeline model described in the main paper for its simplicity and its suggestively. We looked for papers that described errors arising at  decisions points immediately adjacent to stages in our pipeline as well as those in the intersection of multiple decision types.

\vspace{6in}

\onecolumn
  \begin{longtable}{>{\raggedright\arraybackslash}p{3cm}p{14cm}}
    \caption{An expanded collection of examples of errors resulting in mirages along different stages of our analytics pipeline, sorted by the analytical step we believe is responsible for the resulting failure in the final visualization, and colored following \figref{fig:supplement-view}. Just as we highlight in the table in the main paper, this list is not exhaustive. Instead it presents examples of how decision-making at various stages of analysis can damage the credibility or reliability of the messages in charts.}

  \\\hbox{\normalsize{\textbf{CURATING ERRORS}}}&\\ \\
  \normalsize{Error} & \normalsize{Mirage}\\ \hline
   \rowcolor{colora}Forgotten Population or Missing Dataset  & We expect that datasets fully cover or describe phenomena of interest. However, structural, political, and societal biases can result in the over- or under-sampling of populations or problems of importance. This mismatch in coverage can hide crucial concerns about the possible scope of our analyses. \cite{missingdatasets, dignazio2019draft}\\
 \rowcolor{colora-opaque}Geopolitical Boundaries in Question  & Shifting borders and inconsistent standards of ownership can cause geospatial visualizations to be inconsistent. For instance, statistical measures of the United States change significantly depending on whether protectorates and territories are included, or if overseas departments are excluded when calculating measures for France. These issues are more complex when nationstates disagree on the border and extent of their territory, which can cause maps to display significantly different data based on who is viewing the data with what software from what location. \cite{missingdatasets,soeller2016mapwatch}\\

  \\\hbox{\normalsize{\textbf{CURATING + WRANGLING ERRORS}}}&\\ \\
  \normalsize{Error} & \normalsize{Mirage}\\ \hline
   \rowcolor{colora}Missing or  Repeated Records  & We often assume that we have one and only one entry for each datum. However, errors in data entry or integration can result in missing or repeated values that may result in inaccurate aggregates or groupings. \cite{kim2003taxonomy} \\
 \rowcolor{colora-opaque}Outliers  & Many forms of analysis assume data have similar magnitudes and were generated by similar processes. Outliers, whether in the form of erroneous or unexpectedly extreme values, can greatly impact aggregation and discredit the assumptions behind many statistical tests and summaries. \cite{kim2003taxonomy} \\
 \rowcolor{colora}Spelling Mistakes  & Columns of strings are often interpreted as categorical data for the purposes of aggregation. If interpreted in this way, typos or inconsistent spelling and capitalization can create spurious categories, or remove important data from aggregate queries. \cite{wang2019uni}\\
 \rowcolor{colora-opaque}Higher Noise than Effect Size  & We often has access to only a sample of the data, or noisy estimates of an unknown true value. How the uncertainty in these estimates is communicated, and whether or not the viewer is made aware of the relative robustness of the effect in the context of this noise, can affect the resulting confidence viewers have in a particular effect. \cite{hofmann2012graphical,hullman2017imagining}\\
 \rowcolor{colora}Sampling Rate Errors  & Perceived trends in distributions are often subject to the sampling rate at which the underlying data has been curated. This can be problematic as an apparent trend may be an artifact of the sampling rate rather than the data (as is the case visualizations that do not follow the rates suggested by the Nyquist frequency). \cite{kindlmann2014algebraic}\\

  \\\hbox{\normalsize{\textbf{WRANGLING ERRORS}}}&\\ \\
  \normalsize{Error} & \normalsize{Mirage}\\ \hline
   \rowcolor{colorb}Differing Number  of Records by  Group  & Certain summary statistics, including aggregates, are sensitive to sample size. However, the number of records aggregated into a single mark can very dramatically. This mismatch can mask this sensitivity and problematize per-mark comparisons; when combined with differing levels of aggregation, it can result in counter-intuitive results such as Simpson's Paradox. \cite{guo2017you}\\
 \rowcolor{colorb-opaque}Analyst Degrees of Freedom  & Analysts have a tremendous flexibility in how they analyze data. These ``researcher degrees of freedom''~\cite{gelman2013garden} can create conclusions that are highly idiosyncratic to the choices made by the analyst, or in a malicious sense promote ``p-hacking'' where the analyst searches through the parameter space in order to find the best support for a pre-ordained conclusion. A related issue is the ``multiple comparisons problem'' where the analyst makes \emph{so many} choices that at least one configuration, just by happenstance, is likely to appear significant, even if there is no strong signal in the data. \cite{gelman2013garden,pu2018garden,zgraggen2018investigating}\\
 \rowcolor{colorb}Confusing Imputation  & There are many strategies for dealing with missing or incomplete data, including the imputation of new values. How values are imputed, and then how these imputed values are visualized in the context of the rest of the data, can impact how the data are perceived, in the worst case creating spurious trends or group differences that are merely artifacts of how missing values are handled prior to visualization. \cite{song2018s}\\
 \rowcolor{colorb-opaque}Inappropriate/Missing Aggregation  & The size of the dataset is often far larger than what can fit in a particular chart. Aggregation at a particular level of detail is a common technique to reduce the size of the data. However, the choice of aggregation function can lead to differing conclusions based on the underlying distribution of the data. Furthermore, these statistical summaries may fail to capture important features of distribution, such as second-order statistics. Conversely, when a designer fails to apply an aggregation function (or applies one at too low a level of detail), the overplotting, access visual complexity, or reduced discoverability can likewise hide important patterns in the data. \cite{anscombe1973graphs,few2019loom,matejka2017same,salimi2018bias,wall2017warning}\\

  \\\hbox{\normalsize{\textbf{VISUALIZING + WRANGLING ERRORS}}}&\\ \\
  \normalsize{Error} & \normalsize{Mirage}\\ \hline
   \rowcolor{colorc}Outliers Dominate Scale Bounds  & Numeric and color scales are often automatically bound to the extent of the data. If there are a few extrema values, this can result in a renormalization in which much of the data is compressed to a narrow output range, destroying the visual signal of potential trends and variability \cite{correll2016surprise,kindlmann2014algebraic}\\
 \rowcolor{colorc-opaque}Latent Variables Missing  & When communicating information about the relationship between two variables, we assume that we have all relevant data. However, in many cases a latent variable has been excluded from the chart, promoting a spurious or non-causative relationship (for instance, both drowning deaths and ice cream sales are tightly correlated, but are related by a latent variable of external temperature). Even if this variable is present, if the relevant functional dependency is unidentified, the appropriate causal linkage between variables may not be visible in the chart. Similarly, subgroups or subpopulations can exist in datasets that, if not properly separated or identified, can apply universal trends to inappropriate subgroups. \cite{anand2015automatic,wang2019uni}\\
 \rowcolor{colorc}Base Rate Masquerading as Data  & Visualizations comparing rates are often assumed to show the relative rate, rather than the absolute rate. Yet, many displays give prominence to these absolute or base rates (such as population in choropleth maps) rather than encoded variable, causing the reader to understand this base rate as the data rate.  \cite{correll2016surprise}\\
 \rowcolor{colorc-opaque}Concealed  Uncertainty  & Charts that don't indicate that they contain uncertainty risk giving a false impression as well a possible extreme mistrust of the data if the reader realizes the information hasn't been presented clearly. There is also a tendency to incorrectly assume that data is high quality or complete, even without evidence of this veracity. \cite{song2018s, few2019loom, mayrTrust2019, sacha2015role}\\

  \\\hbox{\normalsize{\textbf{VISUALIZING ERRORS}}}&\\ \\
  \normalsize{Error} & \normalsize{Mirage}\\ \hline
   \rowcolor{colorc}Non-sequitur  Visualizations  & Readers expect graphics that appear to be charts to be a mapping between data and image. Visualizations being used as decoration (in which the marks are not related to data) present non-information that might be mistaken for real information. Even if the data are accurate, additional unjustified annotations could produce misleading impressions, such as decorating uncorrelated data with a spurious line of best fit. \cite{correll2017black}\\
 \rowcolor{colorc-opaque}Misunderstand Area as Quantity  & The use of area encoded marks assumes readers will be able to visually compare those areas. Area encoded marks are often misunderstood as encoding length which can cause ambiguity about interpretation of magnitude. \cite{pandey2015deceptive, correll2017black}\\
 \rowcolor{colorc}Non-discriminable Colors  & The use of color as a data-encoding channel presumes the perceptual discriminability of colors. Poorly chosen color palettes, especially when marks are small or cluttered, can result in ambiguity about which marks belong to which color classes. \cite{szafir2017modeling}\\
 \rowcolor{colorc-opaque}Unconventional Scale Directions  & Viewers have certain prior expectations on the direction of scales. For instance, in languages with left-to-right reading orders, time is likewise assumed to move left to right in graphs. Depending on context, dark or opaque colors are perceived as having higher magnitude values than brighter or more transparent colors. Violating these assumptions can cause slower reading times or even the reversal of perceived trends. \cite{correll2017black,pandey2015deceptive,tversky1991cross,schloss2018mapping}\\
 \rowcolor{colorc}Overplotting  & We typically expect to be able to clearly identify individual marks, and expect that one visual mark corresponds to a single value or aggregated value. Yet overlapping marks can hide internal structures in the distribution or disguise potential data quality issues. \cite{correll2018looks,mayorga2013splatterplots,micallef2017towards}\\
 \rowcolor{colorc-opaque}Singularities  & In chart types, such as line series or parallel coordinates plots, many data series can converge into a single point in visual space. Without intervention, viewers can have issues discriminating between which series takes which path after such a singularity. \cite{kindlmann2014algebraic}\\
 \rowcolor{colorc}Inappropriate Semantic Color Scale  & Colors have different effects and semantic associations depending on context (for instance the cultural context of green being associated with money in the United States). Color encodings in charts that violate these assumptions can result in viewers misinterpreting the data: for instance, a viewer might be confused by a map in which the oceans are colored green, and the land colored blue. \cite{lin2013selecting}\\
 \rowcolor{colorc-opaque}Within-the-Bar-Bias  & The filled in area underneath a bar chart does not communicate any information about likelihood. However, viewers often erroneously presume that values inside the visual area of the bar are likelier or more probable than values outside of this region, leading to erroneous or biased conclusions about uncertainty. \cite{correll2014error,newman2012bar}\\
 \rowcolor{colorc}Clipped Outliers  & Charts are often assumed to show the full extent of their input data. A chosen domain might exclude meaningful outliers, causing some trends in the data to be invisible to the reader. \\
 \rowcolor{colorc-opaque}Continuous Marks for Nominal Quantities  & Conventionally readers assume lines indicate continuous quantities and bars indicate discrete quantities. Breaking from this convention, for instance using lines for nominal measures, may cause readers to hallucinate non-existent trends based on ordering.  \cite{mcnuttlinting, zacks1999bars}\\
 \rowcolor{colorc}Modifiable Areal Unit Problem  & Spatial aggregates are often assumed to be presenting their data without bias, yet they are highly dependent on the shapes of the bins defining those aggregates. This can cause readers to misunderstand the trends present in the data. \cite{fotheringham1991modifiable, kindlmann2014algebraic}\\
 \rowcolor{colorc-opaque}Manipulation of  Scales  & The axes and scales of a chart are presumed to straightforwardly represent quantitative information. However, manipulation of these scales (for instance, by flipping them from their commonly assumed directions, truncating or expanding them with respect to the range of the data~\cite{pandey2015deceptive, correll2017black, cleveland1982variables, ritchie2019lie, correll2019truncating}, using non-linear transforms, or employing dual axes~\cite{KindlmannAlgebraicVisPedagogyPDV2016, cairo2015graphics}) can cause viewers to misinterpret the data in a chart, for instance by exaggerating correlation~\cite{cleveland1982variables}, exaggerating effect size~\cite{correll2019truncating,pandey2015deceptive}, or misinterpreting the direction of effects~\cite{pandey2015deceptive}. \cite{cairo2015graphics,correll2017black,correll2019truncating,cleveland1982variables,KindlmannAlgebraicVisPedagogyPDV2016,pandey2015deceptive,ritchie2019lie}\\
 \rowcolor{colorc}Trend in Dual Y-Axis Charts are Arbitrary  & Multiple line series appearing on a common axis are often read as being related through an objective scaling. Yet, when y-axes are superimposed the relative selection of scaling is arbitrary, which can cause readers to misunderstand the magnitudes of relative trends. \cite{KindlmannAlgebraicVisPedagogyPDV2016, cairo2015graphics}\\
 \rowcolor{colorc-opaque}Nominal Choropleth Conflates Color Area with Classed Statistic  & Choropleth maps color spatial regions according to a theme of interest. However, the size of these spatial regions may not correspond well with the actual trend in the data. For instance, U.S. Presidential election maps colored by county can communicate an incorrect impression of which candidate won the popular vote, as many counties with large area have small populations, and vice versa. \cite{gastner2005maps,nusrat2016state}\\
 \rowcolor{colorc}Overwhelming Visual Complexity  & We may assume that there is a benefit to presenting all of the data in all of its complexity. However, visualizations with too much visual complexity can overwhelm or confuse the viewer and hide important trends, as with graph visualization ``hairballs.'' \cite{hofmann2012graphical, greadability}\\

  \\\hbox{\normalsize{\textbf{READING ERRORS}}}&\\ \\
  \normalsize{Error} & \normalsize{Mirage}\\ \hline
   \rowcolor{colord}Reification  & It can be easier to interpret a chart or map as being a literal view of the real world, rather than to understand that it as abstraction at the end of a causal chain of decision-making. That is, as confusing the \emph{map} with the \emph{territory}. This misunderstanding can lead to falsely placed confidence in measures containing flaws or uncertainty: Drucker~\cite{drucker2012humanistic} claims that reification caused by information visualization results in a situation ``as if all critical thought had been precipitously and completely jettisoned.'' \cite{drucker2012humanistic}\\
 \rowcolor{colord-opaque}Assumptions of Causality  & We assume that highly correlated data plotted in the same graph have some important linkage. However, through visual design or arbitrary juxtaposition, viewers can come away with erroneous impressions of relation or causation of unrelated or non-causally linked variables. \cite{xiong2019illusion, few2019loom}\\
 \rowcolor{colord}Base Rate Bias  & Readers assume unexpected values in a visualization are emblematic of reliable differences. However, readers may be unaware of relevant base rates: either the relative likelihood of what is seen as a surprising value or the false discovery rate of the entire analytic process. \cite{correll2016surprise,pu2018garden, zgraggen2018investigating}\\
 \rowcolor{colord-opaque}Inaccessible Charts  & Charts makers often assume that their readers are homogeneous groups. Yet, the way that people read charts is heterogeneous and dependent on perceptual abilities and cognitive backgrounds that can be overlooked by the designer. Insufficient mindfulness of these differences can result in miscommunication. For instance, a viewer with color vision deficiency may interpret two colors as identical when the designer intended them to be separate or a viewer with dyslexia might mistake similarity named points in a annotated scatter plot as denoting the same entity. \cite{lundgard2019Sociotechnical, plaisant2005information, wutanisszafir2019}\\
 \rowcolor{colord}Default Effect  & While default settings in visualization systems are often selected to guide users towards best practices, these defaults can have an outsized impact on the resulting design. This influence can result in mirages: for instance, default color palettes can artificially associate unrelated variables; or default histogram settings can hide important data quality issues. \cite{correll2018looks,few2019loom, hullman2011visualization,shah2006policy}\\
 \rowcolor{colord-opaque}Anchoring Effect  & Initial framings of information tend to guide subsequent judgements. This can cause readers to place undue rhetorical weight on early observations, which may cause them to undervalue or distrust later observations.  \cite{ritchie2019lie, hullman2011visualization}\\
 \rowcolor{colord}Biases in  Interpretation  & Each viewer arrives to a visualization with their own preconceptions, biases, and epistemic frameworks. If these biases are not carefully considered, various cognitive biases such as the backfire effect or confirmation bias can cause viewers to anchor on only the data (or the reading of the data) that supports their preconceived notions, reject data that does not accord with their views, and generally ignore a more holistic picture of the strength of the evidence. \cite{dignazio2019draft, d2016feminist, few2019loom,wall2017warning,valdez2017framework}\\

  \\\hbox{\normalsize{\textbf{READING + WRANGLING ERRORS}}}&\\ \\
  \normalsize{Error} & \normalsize{Mirage}\\ \hline
   \rowcolor{colord}Drill-down Bias  & We assume that the order in which we investigate our data should not impact our conclusions. However, by filtering on less explanatory or relevant variables first, the full scope of the impact of later variables can be hidden. This results in insights that address only small parts of the data, when they might be true of the larger whole. \cite{lee2019avoiding}\\
 \rowcolor{colord-opaque}Cherry Picking  & Filtering and subsetting are meant to be tools to remove irrelevant data, or allow the analyst to focus on a particular area of interest. However, if this filtering is too aggressive, or if the analyst focuses on individual examples rather than the general trend, this cherry-picking can promote erroneous conclusions or biased views of the relationships between variables. Failing to keep the broader dataset in context can also result in the Texas Sharpshooter Fallacy or other forms of HARKing~\cite{cockburn2018hark}. \cite{few2019loom}\\
 \rowcolor{colord}Availability Heuristic  & Examples that are easier to recall are perceived as more typical than they actually are. In a visual analytics context, this could be reflected in analysts recalling outlying instances more easily than values that match the trend, or assuming that the data patterns they encounter most frequently (for instance, in the default or home view of their tool) are more common than they really are in the dataset as a whole. \cite{dimara2016accounting,dimara2018task,few2019loom}\\
  
  \end{longtable}
  \label{table:big-mirage-table}

\end{document}